\documentclass[12pt,a4paper]{article}

\newcommand{\templatetitle}{Procedural Mixture
  Sets} %ENTER PAPER TITLE

\usepackage{authblk, varwidth, etoolbox}\NewCommandCopy{\oldauthor}{\author}\NewCommandCopy{\oldaffil}{\affil}\newcommand{\templateauthors}{}
\renewcommand{\author}[2][2]{%
  \ifdefempty{\templateauthors}
    {\appto\templateauthors{#2}}
    {\appto\templateauthors{, #2}}
    \oldauthor[#1]{#2}%
  }
  \renewcommand{\affil}[2][2]{\oldaffil[#1]{%
      \begin{minipage}[t]{12.5cm}\protect\footnotesize
        #2
      \end{minipage}%
    }}

\author[1,2]{Rommeswinkel, Hendrik}

\affil[1]{Hitotsubashi University, 2-1 Naka, Kunitachi, Tokyo
  186-0004, hendrik.r@r.hit-u.ac.jp}
\affil[2]{Waseda Institute for Advanced Study, 1-104 Totsukamachi, Shinjuku City, Tokyo 169-8050}

\newcommand{\templatesubject}{Economics} %ENTER SUBJECT OF ARTICLE (empty by default)
\newcommand{\templatekeywords}{Decision theory, procedural value,
  decision processes, mixture sets, entropy, reduction of compound
  mixtures, decision times,
  associativity} %ENTER KEYWORDS (empty by default)

\newcommand{\templatethanks}{\thanks{Jaffray Lecture at Risk,
    Uncertainty, and Decision 2024. Many thanks to participants at the
    Society for the Advancement of Economic Theory conference 2019 and
    Risk, Uncertainty, and Decision Conference 2024, seminar
    participants at Erasmus Universiteit Rotterdam and Waseda
    University and Chun-ting Chen, Soo Hong Chew, and Patrick
    DeJarnette for helpful comments. }}

\title{\templatetitle\templatethanks} %
\date{This version: August 2025} %

\usepackage[utf8]{inputenc}
\usepackage{csquotes}
\usepackage[english]{babel}

\usepackage[sc,osf]{mathpazo}
\usepackage{amsthm, amsmath, amssymb}

\usepackage[mathscr]{euscript} %provides \mathscr font
\usepackage{microtype} %fixes some bad scripts
\usepackage{mathtools} %this allows to use \mathclap to remove whitespace due to subscripts and \mathstack for multiline subscripts

\usepackage[backend=biber, style=apa, autocite=plain]{biblatex}
\DeclareLanguageMapping{english}{english-apa}
\usepackage{tikz}
\usepackage{lipsum}
\usepackage{comment}

\usepackage{titling}

\usepackage{sectsty}
\makeatletter\def\@seccntformat#1{\protect\makebox[0pt][r]{\csname the#1\endcsname\hspace{11pt}}}\makeatother
\allsectionsfont{\sc}
\newcommand{\printkeywords}{{\sc Keywords:} \templatekeywords}

\usepackage{booktabs}
\usepackage{array}
\usepackage{float}
\usepackage{multirow}

\usepackage[allcolors=black]{hyperref}
\hypersetup{
    pdftitle={\templatetitle},
    pdfauthor={\templateauthors},
    pdfsubject={\templatesubject},
    pdfkeywords={\templatekeywords},
    bookmarksnumbered=true,
    bookmarksopen=true,
    bookmarksopenlevel=1,
    colorlinks=true,
    pdfstartview=Fit,
    pdfpagemode=UseOutlines,    % this is the option you were lookin for
    pdfpagelayout=TwoPageRight
}

\usepackage{tabularx}
\usepackage{ellipsis} %makes \ldots nicer, must be loaded after hyperref

{
{\theoremstyle{definition}\newtheorem{definition}{Definition}}
{\theoremstyle{remark}\newtheorem{remark}{Remark}}
{\theoremstyle{definition}}
{\theoremstyle{definition}\newtheorem{axiom}{Axiom}}
\newtheorem{theorem}{Theorem}
\newtheorem{corollary}{Corollary}
\newtheorem{proposition}{Proposition}
\newtheorem{lemma}{Lemma}
}
\newcommand*{\myexamplename}{Example}
\newenvironment{example}[1][\myexamplename]{\begin{proof}[#1]}{\end{proof}}
\begin{document}

\maketitle
\begin{abstract}\noindent
  The paper characterizes the \textcite{shannon_mathematical_1948} and
  \textcite{tsallis_possible_1988} entropies in a standard framework
  of decision theory, mixture sets. Procedural mixture sets are
  introduced as a variant of mixture sets in which it is not
  necessarily true that a mixture of two identical elements yields the
  same element. This allows the process of mixing itself to have an
  intrinsic value. The paper proves the surprising result that simply
  imposing the standard axioms of von Neumann-Morgenstern on
  preferences on a procedural mixture set yields the entropy as a
  representation of procedural value. An application of the theorem to
  decision processes and the relation between choice probabilities and
  decision times elucidates the difficulty of extending the
  drift-diffusion model to multi-alternative choice.

    \noindent\printkeywords
\end{abstract}
\section{Introduction}
Economic theory frequently utilizes information measures.\footnote{The
  entropy is used to model inequality
  \parencite{shorrocks_class_1980,theil_economics_1967}, segregation
  \parencite{frankel_measuring_2011}, the utility of gambling
  \parencite{luce_utility_2008-1,luce_utility_2008}, diversity
  \parencite{nehring_diversity_2009}, consumer demand
  \parencite{theil_information_1965}, freedom of choice
  \parencite{suppes_nature_1996}, market concentration
  \parencite{hennessy_when_2007,herfindahl_concentration_1950,hirschman_national_1980},
  and information costs
  \parencite{sims_implications_2003,caplin_rationally_2017}.} The
present paper characterizes the entropy information measures by
\textcite{shannon_mathematical_1948} and
\textcite{tsallis_possible_1988} in a simple algebraic framework.
Concretely, we employ variations of the mixture sets and classical
expected utility axioms of
\textcite{von_neumann_theory_1944,herstein_axiomatic_1953}.

In standard mixture sets, from the reduction of compound mixtures
axiom and the independence axiom follows betweenness
\parencite{chew_axiomatic_1989}: if $a \succsim b$, then $a \succsim \mu a \oplus (1-\mu) b \succsim
b$. The entropy representation characterized in the present paper
violates betweenness. This is natural since the entropy is a measure
of statistical dispersion, not of average value. This leaves two
avenues for characterization: changing the reduction of compound
mixtures or the independence axiom. We focus on the former method and show
that the result can be directly applied to the latter method.

In the first main result we change the reduction of compound mixtures
axiom of mixture sets into an associativity condition. Associativity
is a new axiom that states that the order of mixing
does not matter but unlike reduction of compound mixtures allows for
$\mu a \oplus (1-\mu) a \not\sim a$. We call sets that fulfill these assumptions
procedural mixture sets because not only the result $a$ matters, but
also whether a mixture operation was performed or not. Surprisingly,
the change from a mixture set to a procedural mixture set is
all that is needed to obtain an entropy representation: in a procedural
mixture set, the expected utility axioms --- weak order, continuity, and
independence characterize the Shannon and Tsallis entropies.

We apply our model of procedural value to axiomatic characterizations
of decision times
\parencite{koida_multiattribute_2017,echenique_response_2017,baldassi_behavioral_2020,fudenberg_testing_2020}
in decision processes with multiple alternatives. We show that
decision processes in which choice probabilities follow the
\textcite{luce_individual_1959} model of stochastic choice form a
procedural mixture set. Consider a relation $\succsim$ on decision processes
with the interpretation of ``takes as least as long as'' which
represents a procedural value (time) as opposed to a consequentialist
value (utility). Betweenness is naturally violated if choice is
non-instantaneous because even the decision between two identical
options may take some additional time. The weak order, continuity, and
independence axioms naturally apply to this relation: in particular,
the independence axiom can be understood as stating that decision
times are increasing in the decision times of sub-decisions. The
longer a sub-decision takes, the longer --ceteris paribus-- the overall
decision should take. Applying the main result to decision processes
thus yields a testable empirical prediction: if for a decision domain
the \textcite{luce_individual_1959} model holds and decision times are
a function of decision probabilities that is increasing in the
decision times of subdecisions, then decision times can be represented
by an entropy. \textcite{hick_rate_1952} first suggested that decision
durations can be modeled by the \textcite{shannon_mathematical_1948}
entropy and thus we call the resulting model of decision processes the
Luce-Hick model, which we characterize axiomatically. We use this result to
explain the well-known difficulty of finding multi-alternative
extensions for the Diffusion Drift model
\textcite{ratcliff_theory_1978,ratcliff_diffusion_2016}.

In the second main result we restrict the independence axiom to
mixtures with disjoint support but maintain the mixture set
assumptions introduced by \textcite{herstein_axiomatic_1953}. By
embedding a procedural mixture set into the mixture set, we again
obtain the Shannon and Tsallis entropy representation for a subset of
the relation. We apply this result to nested stochastic choice
\parencite{kovac_behavioral_2022}. In nested stochastic choice Luce's
IIA axiom holds among subsets of alternatives. This allows us to
generate Hick's law type predictions for decision times when some
pairs of alternatives are similar and others are dissimilar to each
other.

The paper proceeds as follows. An introductory example highlights the
main application to decision processes in Section \ref{sec:example}.
The axioms and the representation theorem for procedural mixtures are
presented in Section \ref{sec:axioms}. Section
\ref{sec:comparativeStatics} provides comparative statics results and
interprets the parameters of the model. Section
\ref{sec:disjointIndependence} discusses the representation results
that maintain reduction of compound mixtures but weaken independence.
The relation to the literature is given in Section
\ref{sec:literature}. Section \ref{sec:discussion} concludes.

\section{Introductory Example}\label{sec:example}
Suppose an analyst observes a sample of choices by a decision maker
choosing between various items. This choice is not instantaneous; the
decision maker undergoes some thought process until a decision is
made. The duration of this thought process will generally depend on
how difficult the decision is. The literature on decision processes
\parencite{usher_dynamics_2013, bogacz_physics_2006} has established
that the more uncertain ex ante the choice between two alternatives
is, the longer a decision takes. Thus, when two alternatives are
equally likely to be chosen, then the decision process takes longer
than when one alternative is much more likely to be chosen than the
other.\footnote{In the experimental literature, the decision time
  commonly refers to the average decision time of a sample of subjects
  and the choice probabilities refer to the relative frequency of
  choice.} The drift diffusion model \parencite{ratcliff_theory_1978}
captures this basic empirical fact but there is no agreed-upon
generalization to multiple options
\parencite{ratcliff_diffusion_2016}. The generalization to multiple
options is complicated by the presence of similarity, attraction, and
compromise effects
\parencite{roe_multialternative_2001,rieskamp_extending_2006} in
stochastic choice. Generally, the availability of additional
alternatives may influence the relative choice probabilities of two
alternatives, complicating the extension of the relation between
choice probabilities and decision times from two options to multiple
options. However, as we will show in Section \ref{sec:axioms}, in the
absence of the aforementioned effects, it turns out that plausible
assumptions lead to a very restrictive functional form for the
relation between choice probabilities and decision times.

Consider the classical example of a decision maker who faces a choice
between various methods of transportation to travel to another city.
An analyst records the choice probabilities and decision times shown
in Figure \ref{fig:decisionData}. On the left side, the table lists
decision problems with choice probabilities and on the right side,
decision times. We will call the combination of a decision problem
(i.e., the list of available alternatives) and choice probabilities a
{\em decision process}. From their decision times we can derive a weak
order over decision processes. If $a$ and $b$ are decision processes,
then $a \succsim b$ is interpreted as $a$ takes at least as long as $b$. Such
a relation will form the primitive of our model.

\begin{figure}[ht]
  % These decision times have been obtained by parameters r=1, q=1/ln(2), U(x)=.5
  \centering
  \begin{tabular}{ccc|c}
    Option 1 & Option 2 & Option 3 & Duration \\
    \hline
    Airplane & {\sc Bus} & {\sc Car} & \multirow{2}{*}{1.9s}\\
    20\% & 20\%& 60\% & \\
    {\sc Airplane} & {\sc Bus} & - & \multirow{2}{*}{1.5s}\\
    50\% & 50\%& & \\
    {\sc Bus} & {\sc Bus} & - & \multirow{2}{*}{1.5s}\\
    50\% & 50\%& & \\
    {\sc Airplane} & {\sc Car} & - & \multirow{2}{*}{1.3s}\\
    25\% &  75\% & & \\
    {\sc Bus} & {\sc Car} & - & \multirow{2}{*}{1.3s}\\
    25\% &  75\% & & \\
    {\sc Airplane} & {\sc Train} & - & \multirow{2}{*}{1.2s}\\
    20\% & 80\% & & \\
    {\sc Bus} & - & - & \multirow{2}{*}{.5s}\\
    100\% & & & \\
    {\sc Train} & - & - & \multirow{2}{*}{.5s}\\
    100\% & & &
  \end{tabular}
  \caption{Choice Probability and Decision Time Data}
  \label{fig:decisionData}
\end{figure}

The listed decision processes exhibit two important features that
motivate the formal model presented in the next section:

First, there are decision problems which list the same alternative
more than once. For example, there is a trivial decision between
taking a bus and taking a bus. Such a trivial decision may still
involve a reaction time or may even induce the decision maker to
engage in a search for differences between the two alternatives.
Treating decision processes simply as probability distributions over
what alternative is chosen would conflate such a decision process with
a process that involves no choice. We will later make this precise by
defining that the decision processes in Figure \ref{fig:decisionData}
lack {\em reducibility}.

Second, the relative choice probabilities of two alternatives do not
depend on other alternatives available: the airplane and bus are
equally likely to be chosen when compared with each other and this
does not change when the car is available. This property is an
instance of the independence of irrelevant alternatives (IIA) axiom of
\textcite{luce_individual_1959}. The probabilities of the three option
decision problem in Figure \ref{fig:decisionData} are such that
conditioning on any subset yields the corresponding binary choice
probabilities. Decisions can therefore be seen as composed of binary
sub-decisions. The IIA axiom guarantees that the order in which we
compose such binary sub-decisions does not matter. We will later make
this precise by defining that the decision processes in Figure
\ref{fig:decisionData} fulfill {\em associativity}.

The above two observations motivate us to study {\em procedural
  mixtures} denoted by $\mu a \oplus (1-\mu) b$ that fulfill associativity but,
unlike standard mixture sets, do not fulfill reducibility. If $a$ and
$b$ are decision processes from Figure \ref{fig:decisionData}, then $\mu
a \oplus (1-\mu) b$ is the decision process which assigns each option in $a$
and $b$ its original choice probability multiplied with $\mu$ and $1-\mu$,
respectively. Crucially, if $a$ and $b$ contain identical or repeated
options, these are not reduced to a single instance with the
probability being the sum. They remain separately listed as the
decision maker may take additional time to discern between the two
identical options.

For example, denoting by $[\textsc{Bus}]$ the trivial decision problem
that only offers to ride the bus, we can denote by $1/2 [\textsc{Bus}]
\oplus 1/2 [\textsc{Bus}]$ the decision problem in which the (of course,
meaningless) choice between a bus and a bus is made. Importantly,
unlike in mixture sets, $[\textsc{Bus}] \neq 1/2 [\textsc{Bus}] \oplus 1/2
[\textsc{Bus}]$ as required by our first observation. The choice
between the airplane, the bus, and the car can be written as $1/5
[\textsc{Airplane}] \oplus 4/5 (1/4 [\textsc{Bus}] \oplus 3/4 [\textsc{Car}])$
but also as $2/5 (1/2 [\textsc{Airplane}] \oplus 1/2 [\textsc{Bus}]) \oplus 3/5
[\textsc{Car}]$. Thus, every decision process can be written as an
arbitrary composition of binary subdecisions. Luce's IIA axiom
guarantees that the mixture weights correspond to the true choice
probabilities of the subdecision problems and that our decision
processes can consistently be disaggregated into binary choices. This
corresponds to our second observation discussed above.

The procedural mixture notation allows us to write a decision process
in terms of its {\em sub-decision processes}. For example, $(1/4
[\textsc{Bus}] \oplus 3/4 [\textsc{Car}])$ is a sub-decision process of
$1/5 [\textsc{Airplane}] \oplus 4/5 (1/4 [\textsc{Bus}] \oplus 3/4
[\textsc{Car}])$. In Figure \ref{fig:decisionData} we observe that the
decision time of a choice process is increasing in the decision time
of any sub-decision process. For example, the decision process between
the airplane, bus and the car takes longer than the decision process
between the airplane and the train just as the decision between the
bus and the car takes longer than being assigned the train with
certainty. This corresponds to the independence axiom
\parencite{von_neumann_theory_1944,herstein_axiomatic_1953}, being
applied to a procedural mixture set. According to the independence
axiom, $[\textsc{Airplane}] \succsim [\textsc{Bus}]$, i.e.,
$[\textsc{Airplane}]$ takes at least as long as $[\textsc{Bus}]$, if
and only if $1/2 [\textsc{Airplane}] \oplus 1/2 [\textsc{Car}] \succsim 1/2
[\textsc{Bus}] \oplus 1/2 [\textsc{Car}]$. In Section \ref{sec:axioms} we
show that this assumption (together with continuity of the relation on
a rich\footnote{Technically, the result requires arbitrary decision
  probabilities to be available for all alternatives to fulfill
  richness. In practice, sufficiently fine probabilities can be
  generated from quality variations, probabilities of actually
  receiving an item, etc..} set of decision probabilities) allows us
to characterize a sharp functional form for the average decision time;
the Tsallis and Shannon entropies of the choice probabilities.

It is often perceived as a weakness of the Luce model that adding an
(almost) identical option to a decision increases the (combined)
probability of that option \parencite[e.g., ][]{debreu_review_1960}.
This is because all alternatives are treated symmetrically and the
qualitative differences do not matter. Analogously, in our model, the
(meaningless) decision between two buses takes as much time as the
decision between an airplane and a bus. We address this issue with a
second set of results in Section \ref{sec:disjointIndependence} where
we allow for the decision between the two buses and the decision
between a bus and an airplane to be treated differently and show that
in a mixture set, a suitably weakening of the independence axiom
yields the same representation.

\section{Procedural Mixtures}\label{sec:axioms}
For ease of comparison, we first recapitulate the axioms of
\textcite{herstein_axiomatic_1953} which for a mixture set $\langle
\mathscr{M}, \oplus \rangle$ where $\oplus$ is a mixture operator $\oplus: \mathscr{S} \times
\mathscr{S} \times [0,1] \rightarrow \mathscr{S}$ are given as follows:
\begin{align}
    &1a \oplus (1-1)b = a,\\
    &\mu a \oplus (1-\mu) b = (1-\mu)b \oplus \mu a,\\
    &\lambda[\mu a \oplus (1-\mu) b] \oplus (1-\lambda) b = (\lambda \mu) a \oplus (1-\lambda \mu) b
\end{align}
where each axiom holds for all $a,b\in \mathscr{M}$ and all $\mu, \lambda \in
[0,1]$. We may call these axioms, respectively, connectedness,
commutativity, and reduction of compound mixtures. $\lambda$ and $\mu$ are the
mixture weights. In the context of our example, these are
probabilities and we will use these terms interchangeably. An element
$a$ is called an {\em outcome} if there do not exist distinct $b$ and
$c$ such that $a = \mu b \oplus (1-\mu) c$ for some $\mu \in (0,1)$.

The reduction of compound mixtures axiom is implied\footnote{Notably,
  associativity is not implied by reduction of compound mixtures as
  Example 1 in \textcite{mongin_note_2001} shows.} by two economically
distinct properties, associativity,
\begin{align}
  & (1-\lambda)[\frac{\mu}{1-\lambda} a \oplus \frac{1-\lambda-\mu}{1-\lambda} b] \oplus \lambda c \nonumber\\
    = &
    \mu a \oplus (1-\mu)\left[ \frac{1-\lambda-\mu}{1-\mu} b \oplus \frac{\lambda}{1-\mu} c\right]
\end{align}
and reducibility:
\begin{align}
  & \mu a \oplus (1-\mu) a = a.
\end{align}
for all $a,b\in \mathscr{M}$ and all $\mu \in [0,1]$ and $\lambda \in [0,1)$.
Associativity states that the order of mixing does not matter.
Reducibility expresses that the mixing itself is irrelevant.

\begin{example}
  The classical example of a mixture set is a set of lotteries on a
  set of alternatives $\mathscr{X}$. These can be formalized as
  $\Delta X \equiv \{p:
  \mathscr{X} \rightarrow [0,1] | \sum_{x \in \mathscr{X}}p(x) = 1\}$ with a mixture
  operation fulfilling $(\alpha p \oplus (1-\alpha) q) (x) = \alpha p(x) + (1-\alpha) q(x)$ for
  all $x \in \mathscr{X}$. Notice that in our example, a decision
  process in which a decision maker chooses bus with probability $1/2$
  or another bus is not the same as the trivial decision process in
  which the decision maker does not make a choice and receives a bus
  with certainty. Lotteries and mixture sets would not reflect this
  since $\frac{1}{2} b \oplus \frac{1}{2} b = b $. In procedural mixture
  sets we therefore remove the reducibility axiom to allow the
  procedural mixture set to distinguish between the decision process
  involving a choice, $\frac{1}{2} b \oplus \frac{1}{2} b$ and the trivial
  decision process involving no choice, $b$. Thus, the mixture
  operation $\mu a \oplus (1-\mu) b$ means in this context that the decision
  maker makes a time-consuming decision with probabilities $\mu$ and
  $1-\mu$ between alternatives $a$ and $b$ and that this decision is
  time-consuming even if the alternatives are effectively identical.
\end{example}

\textcite{fishburn_foundations_1982} generalized mixture sets by
replacing the identity $=$ with an equivalence relation in the mixture
axioms. We now remove the axiom of reducibility and perform a
generalization analogous to \textcite{fishburn_foundations_1982}.

\begin{definition}[Procedural Mixture Set]
  A procedural mixture set $\langle \mathscr{S},\oplus,\approx \rangle$ is a set
  $\mathscr{S}$ endowed with a mixture operator $\oplus: \mathscr{S} \times
  \mathscr{S} \times [0,1] \rightarrow \mathscr{S}$ and an equivalence relation $\approx$
  which fulfills for all $a,b,c \in \mathscr{S}$ and all $\mu \in [0,1]$, $\lambda
  \in [0,1)$:
  \begin{align}
   &1a \oplus (1-1)b \approx a,\label{eq:connectedness}\\
   &\mu a \oplus (1-\mu) b \approx (1-\mu)b \oplus \mu a,\label{eq:commutativity}\\
   & (1-\lambda)\left[\frac{\mu}{1-\lambda} a \oplus \frac{1-\lambda-\mu}{1-\lambda} b\right] \oplus \lambda c \nonumber\\
   & \approx
    \mu a \oplus (1-\mu)\left[ \frac{1-\lambda-\mu}{1-\mu} b \oplus \frac{\lambda}{1-\mu} c\right]
     \label{eq:associativity}
\end{align}
\end{definition}
Connectedness and commutativity remain unchanged. Reduction of
compound mixtures is replaced by associativity.

\begin{example}
	The data structure consisting of entries in the form of rows in Figure
	\ref{fig:decisionData} is a procedural mixture set given our
	assumptions about choice probabilities and equivalent decision times discussed
	in Section \ref{sec:example}. To see this, we now specify the set $\mathscr{S}$, the
	operation $\oplus$, the equivalence relation $\approx$, and the resulting equivalence
	classes $\mathscr{S}/\approx$.

	Let $\mathscr{X}$ be a set of outcomes. A decision process with $n$
  outcomes is an element of $\mathscr{S}_n \equiv \mathscr{X}^n\times
  \Delta\{1,\ldots,n\}$, i.e., a list of $n$ (possibly identical) outcomes
  and a probability distribution over the $n$ outcomes. Denote an
  element of $\mathscr{S}_n$ by $(x_1,p_1;\ldots;x_n,p_n)$. Let
  $\mathscr{S} = \bigcup_{n=1}^{\infty} \mathscr{S}_n$. We define the
  operator $\oplus$ as follows:
	\begin{align}
		\mu a \oplus (1-\mu) b =
		\begin{cases}
			a,                                                                                      & \mu=1        \\
			b,                                                                                      & \mu=0        \\
			(x^a_1, \mu p_1^a;\ldots;x^a_{n^a}, \mu p_{n^a}^a;x^b_1,(1-\mu)p_1^b;\ldots;x^b_{n^b}), & \text{else.}
		\end{cases}
	\end{align}

	We further define the equivalence relation $\approx$ such that $a \approx b$ if
  $b$ can be obtained from $a$ by removing or adding any number of
  zero probability outcomes and permuting the combinations of outcomes
  and probabilities.

	The crucial difference to a standard lottery space is that whenever
	two outcomes appear more than once in the decision process, these
	are not treated as the same outcome with the probability being the
	sum of the two times they appear.
\end{example}

We are interested in binary relations on procedural mixture sets. A
function $U: \mathscr{S} \rightarrow \mathbb{R}$ is called a {\em representation} of $\succsim$
if $a \succsim b$ if and only if $U(a) \geq U(b)$. In particular, we are
interested in the following representation of binary relations on
procedural mixture sets.
\begin{definition}[Mixture Entropy]
  A binary relation $\succsim$ on $\mathscr{S}$ is a mixture entropy value if
  there exists a function $U:\mathscr{S}\rightarrow \mathbb{R}$ called a mixture entropy
  and parameters $q\in \mathbb{R}, r\in \mathbb{R}_{++}$ such that $U$ represents $\succsim$ and
  for all $a,b \in \mathscr{S}$ and $\mu \in [0,1]$,
  \begin{align}
    U(\mu a \oplus (1-\mu)b) =& \mu^{r} U(a) + (1-\mu)^r U(b) + q \cdot H_r(\mu)\nonumber\\
    H_r(\mu) =&
              \begin{cases}
                -\mu \ln \mu - (1-\mu)\ln (1-\mu), & r=1\\
                \frac{1-\mu^r -(1-\mu)^r}{r-1}. & r\neq 1
              \end{cases}\label{eq:entropyMixtureValue}
  \end{align}
  where $0 \ln 0 \equiv 0$ here and thereafter.
\end{definition}

\begin{example}
  If $\succsim$ represents the decision times, then the decision times must
  be a monotone transformation $T$ of the representation $U$. For
  simplicity, assume for the moment that $T$ is the identity function,
  i.e., $T(u)=u$.\footnote{In Section \ref{sec:comparativeStatics}, we
    will see in more detail how the parameters affect decision times
    when $T$ is arbitrary.} For such a linear $T$, the parameters that
  lead to the decision times in Figure \ref{fig:decisionData} are
  characterized as follows: First, each trivial decision (where only
  one option is available) takes a {reaction time} of 0.5 seconds.
  This can be seen as the non-decision component of a response time
  \parencite{luce_response_1986}. Second, the most difficult binary
  decision (in which the choice probabilities are equal) takes 1.5
  seconds. Let the {deliberation time} of this decision process be the
  decision time minus the reaction time, i.e. 1 second. Third, given
  any decision process, if we replace every final option by repeating
  the exact same decision process, then the deliberation time doubles.
  For example, four options that are equally likely to be chosen take
  twice the deliberation time, i.e., 2 seconds, as two equally likely
  options. These assumptions together determine the parameters of the
  mixture entropy value as $r=1$ and $k=1/\ln 2$. In the remainder of
  this section, we provide a set of simple axioms that characterize
  when the decision times are an increasing transformation of
  \eqref{eq:entropyMixtureValue}.
\end{example}

Let $\succsim$ be a relation on a procedural mixture set $\mathscr{S}$. We
use the symbols $\sim$ and $\succ$ to denote the symmetric and asymmetric
parts of $\succsim$. We assume the following classical axioms:
\begin{axiom}[Weak Order]\label{axiom:weakOrder}
$\succsim$ is complete and transitive.
\end{axiom}
A weak order is {\em nontrivial} if for some $a,b$, $a \succsim b$ but not $b
\succsim a$.
\begin{axiom}[Continuity]\label{axiom:continuity}
  For any $a,b,c \in \mathscr{S}$, the sets $\{\mu|\mu a \oplus
  (1-\mu)b \succsim c\}$ and $\{\mu|c\succsim \mu a \oplus (1-\mu)b \}$ are closed.
\end{axiom}
\begin{axiom}[Independence]\label{axiom:independence}
  If $a,a',b\in \mathscr{S}$, $\mu\in (0,1)$ then $a\succsim a' \Leftrightarrow \mu a \oplus (1-\mu) b \succsim
  \mu a' \oplus (1-\mu) b$.
\end{axiom}
Our independence axiom needs to be slightly stronger than that of
\textcite{herstein_axiomatic_1953}. Reducibility allows them to generate our
third axiom from a weaker assumption requiring only indifferences.
\begin{example}
  In the context of our example, independence means that the decision
  duration of a decision process is increasing in the duration of
  every sub-decision process. In a process that can be written as $\mu a
  \oplus (1-\mu) b$, the greater the decision time of $a$, the greater the
  decision time of $\mu a \oplus (1-\mu) b$. Specifically, consider the
  comparison of decision times of the decision process between an
  airplane and a car and between a bus and a car. These only differ on
  the sub-decision process in case a car is not chosen. If the
  reaction time of the trivial decision process offering the airplane
  is just as long as the reaction time of being offered the bus,
  independence requires that also the decision process of the choice
  between an airplane and a car takes just as long as the decision
  process of a bus and a car.
\end{example}

\begin{theorem}\label{thm:ProceduralMixture}
  Let $\succsim$ be a binary relation on a procedural mixture set $\langle
  \mathscr{S}, \oplus ,\sim \rangle$. Then the following two statements are
  equivalent.
  \begin{enumerate}
  \item $\succsim$ fulfills axioms
    \ref{axiom:weakOrder}-\ref{axiom:independence}.
  \item $\succsim$ is an entropy mixture value.
  \end{enumerate}
  If $U^1$ and $U^2$ are entropy mixture value representations of the
  same nontrivial weak order, then $r^1=r^2$, and if $r^1=1$, then
  $U^1 = \phi U^2 + \psi$ and $q^1 = \phi q^2$ and if $r^1\neq1$, then $U^1 = \phi
  U^2 + \psi$ and $q^1= \phi q^2 + \psi$ where $\phi\in \mathbb{R}_+$ and $\psi \in \mathbb{R}$.
\end{theorem}
We have characterized two possible representations. Either we obtain
the expected entropy mixture value of the mixed elements plus the
\textcite{shannon_mathematical_1948} entropy. Alternatively, we obtain
the expected entropy mixture value under power-form probability
distortions plus the \textcite{tsallis_possible_1988} entropy. We
delay the interpretation of the parameters of the model until
discussing their comparative statics in Section
\ref{sec:comparativeStatics}. We first show that the characterization
provides a theoretical foundation for the difficulty \parencite[see
][]{ratcliff_diffusion_2016} of finding plausible extensions of the
drift-diffusion model of decision times.

\begin{example}
  It is noteworthy that we have only characterized a representation of
  the decision times and not the exact decision times. Thus, the
  actual decision times can be any increasing transformation of $U$.
  Therefore, the representation with $r=1$, $q>0$ and $U(x)=0$ for
  trivial decision processes $x$ is consistent with any model of {\em
    binary} decision processes in which the decision duration of $\mu x
  \oplus (1-\mu) y$ is strictly increasing in $\min(\mu, 1-\mu)$. This is the
  case for the drift-diffusion model of
  \textcite{ratcliff_theory_1978}.
  To obtain the drift-diffusion model drift, the monotone
  transformation $T: \mathbb{R} \rightarrow \mathbb{R}$ would be
  \begin{align}
    \label{eq:DriftDiffusionTime}
    T = & f \circ (H_1)^{-1} \\
    f(\mu) = & k \cdot \frac{1-\mu - \mu}{\ln(1-\mu) - \ln(\mu)} \\
    (H_1)^{-1}(u) = & \min\{\mu \in [0,1]: H_1(\mu) = u\}.
  \end{align}
  In other words, $T$ first recovers the probability $\mu \leq 1-\mu$ from
  the entropy $H_1(\mu)$ and then applies the formula for the average
  decision time in a drift-diffusion model. This seems like forcing
  the issue but reveals the reason for the difficulty of finding
  extensions of the drift-diffusion model to multiple options: Suppose
  this extension fulfills Axioms 1-3 and agrees with the
  drift-diffusion model on binary choices. Since the decision times of
  choices among two and three alternatives overlap, the awkward form
  of $T$ also applies to three-element choices and for such choices
  $(H_1)^{-1}$ does not recover a choice probability.

  Extensions of the drift-diffusion model therefore face the following
  tradeoff: (1) they may give up on decision times of a process being
  continuously increasing in the decision times of sub-decision
  processes or (2) they may give up on choice probabilities fulfilling
  Luce's IIA axiom or (3) they accept the awkward form of $T$ and try
  to find a stochastic process and boundary conditions that generate
  decision times that can be represented by an entropy.
\end{example}

\begin{remark}
  The multi-mixture representations follow from compounding in a
  straightforward way, e.g.:
\begin{align}
    &U\left(\alpha_1 a_1 \oplus (1-\alpha_1) \left(\frac{\alpha_2}{1-\alpha_1} a_2 \oplus \frac{1-\alpha_1-\alpha_2}{1-\alpha_1} \left( \frac{\alpha_3}{1-\alpha_1-\alpha_2} a_3 \oplus \ldots \right)\right)\right)\nonumber\\
  =& \sum_{i} (\alpha_i)^r U(a_i) + q \cdot H_r(\alpha_1,\ldots) \\
\end{align}
where the mixtures must be finite if $r \leq 1$ and
\begin{align}
  H_r(\alpha_1,\ldots) \equiv &
  \begin{cases}
    - \sum_i \alpha_i \ln \alpha_i, & r=1 \\
    \frac{1-\sum_i (\alpha_i)^r}{r-1} & r\neq1.
  \end{cases}
\end{align}
\end{remark}

In many contexts, the \textcite{shannon_mathematical_1948} measure is the
standard measure of entropy. The use of the Shannon
entropy in the previous representation entails the following property.
\begin{axiom}[Mixture Cancellation]\label{axiom:mixtureCancellation}
  For all $a,a',b,b' \in \mathscr{S}$ and $\mu,\lambda \in (0,1)$,
  \begin{align}
    & (\mu + \lambda) \left(\frac{\mu}{\mu + \lambda} a \oplus \frac{\lambda}{\mu + \lambda} a  \right) \oplus (1-\mu-\lambda) b \nonumber\\
    \sim &
    (\mu + \lambda) \left(\frac{\mu}{\mu + \lambda} a' \oplus \frac{\lambda}{\mu + \lambda} a' \right) \oplus (1-\mu-\lambda) b'\label{eq:rEqualsOnePlus}\\
    \Leftrightarrow\quad
    & (\mu + \lambda) a \oplus (1-\mu-\lambda) b \nonumber\\
    \sim &
           (\mu + \lambda) a' \oplus (1-\mu-\lambda) b'.\label{eq:rEqualsOne}
  \end{align}
\end{axiom}
It is straightforward to apply mixture cancellation to Theorem
\ref{thm:ProceduralMixture} to obtain the following corollary.
\begin{corollary}\label{rem:rEqualsOne}
  Let $\succsim$ be a binary relation on a procedural mixture set $\langle \mathscr{S}, \oplus ,\sim \rangle$. Then
  the following two statements are equivalent.
\begin{enumerate}
\item $\succsim$ fulfills axioms \ref{axiom:weakOrder}-\ref{axiom:mixtureCancellation}.
\item $\succsim$ is an entropy mixture value with $r=1$.
\end{enumerate}
\end{corollary}
\begin{example}
  In the context of decision processes mixture cancellation has a
  straightforward interpretation: Let $a$ be a more difficult decision than $a'$
  and $b'$ be more difficult than $b$ exactly such that \eqref{eq:rEqualsOne}
  holds. The decision processes in \eqref{eq:rEqualsOnePlus} can be understood
  as identical to the ones in \eqref{eq:rEqualsOne} except that
  before\footnote{By associativity, the additional decision could also be
    performed after $a$ or $a'$.} the sub-decisions $a$ and $a'$ an additional
  decision with relative probability $\frac{\mu}{\mu + \lambda}$ is performed. The
  condition thus says that the different deliberation times of $a$ and $a'$ do
  not ``interact'' with the additional deliberation time from adding an
  additional decision. An example of such choices are given by the domain of
  choices that follow Hick's law \parencite{hick_rate_1952}. Hick observed that
  with remarkable precision the response time to press a button in response to a
  signal increases logarithmically in the number of buttons, similar to how the
  Shannon entropy of uniform variables increases logarithmically in the number
  of outcomes. This suggests that $r=1$ and $T$ being the identity are suitable
  to model exact decision times for choices where Hick's law applies.

  However, this might not be the case if the decision makers become increasingly
  constrained in their decision making capability as the number of options
  increases. In this case, we would expect that if $a$ takes longer than $a'$,
  then
  \begin{align}
    \label{eq:exampleREqualsOne}
    & (\mu + \lambda) a \oplus (1-\mu-\lambda) b \nonumber\\
    \quad \sim \quad &
        (\mu + \lambda) a' \oplus (1-\mu-\lambda) b' \nonumber\\
    \Rightarrow \quad \quad &
    \mu \left(\frac{\mu}{\mu + \lambda} a \oplus \frac{\lambda}{\mu + \lambda} a  \right) \oplus (1-\mu-\lambda) b \nonumber\\
    \succ \quad &
        \mu \left(\frac{\mu}{\mu + \lambda} a' \oplus \frac{\lambda}{\mu + \lambda} a' \right) \oplus (1-\mu-\lambda) b'.
  \end{align}
  That is, the additional decision with relative probability $\mu/(\mu +
  \lambda)$ interacts with the decision processes $a$ and $a'$ such that
  decision times increase more when this decision precedes the more
  complicated decision process $a$. As we will see in the comparative
  statics presented in Section \ref{sec:comparativeStatics}, such
  behavior is closely linked to the parameter $r$.
\end{example}

We end this section with a corollary that applies the main theorem to
stochastic choice models and that makes some of the informal
discussion of the example application precise. Given the close link
between Luce's IIA of decision probabilities and associativity, it is
natural to simultaneously characterize the Luce model choice
probabilities and entropy mixture decision times.

Let $\mathscr{X}$ be a set of alternatives and $\mathscr{C}$ be the
set of finite subsets of $\mathscr{X}$. A stochastic choice function
is a function $p: \mathscr{C} \times \mathscr{X} \rightarrow [0,1]$ such that for all
$C \in \mathscr{C}$, $x \not\in C$, $p(x,C)=0$ and $\sum_{x \in C}p(x,C)=1$.
For a more convenient notation, we often write $p_C(x) \equiv p(x,C)$.
For any $C \subseteq D \in \mathscr{C}$, we further define $p_D(C) = \sum_{x \in
  C}p_D(x)$. A decision time $\tau: \mathscr{C} \rightarrow \mathbb{R}_{+}$ is a function
that tells us for every finite subset of alternatives how long it
takes (on average) to make a decision.

We introduce the following joint model\footnote{This model differs
  from decision processes employed for example by
  \textcite{alos-ferrer_time_2021} because for each opportunity set
  only (average) decision times across all options are known instead
  of full probability distributions conditional on the chosen
  element.} of choice probabilities and decision times:
\begin{definition}[Luce-Hick Model]
  A stochastic choice function $p$ and a decision time $\tau$ form a Luce-Hick model if
  \begin{enumerate}
  \item there exists a function $v: \mathscr{X} \rightarrow \mathbb{R}$ such that for all
    $C \in \mathscr{C}$ and $x \in C$,
    \begin{align}
      \label{eq:luceModel}
      p_C(x) = \frac{\exp(v(x))}{\sum_{y \in C} \exp(v(y))} \text{, and}
    \end{align}
  \item there exists a continuous, strictly monotone function $T$ and $r \in
    \mathbb{R}_{++}$ such that for all $C \in \mathscr{C}$,
    \begin{align}
      \label{eq:hickModel}
      T \circ \tau (C)
      = &
          \begin{cases}
            \frac{ 1  - \sum_{x \in C } p_C(x)^r}{r-1} & r \neq 1 \\
            \sum_{x \in C } -p_C(x) \ln p_C(x) & r=1
          \end{cases}
    \end{align}
    and $\tau(\{x\}) = \tau(\{y\}) = T^{-1}(0)$ for all $x,y \in \mathscr{X}$.
  \end{enumerate}
\end{definition}
That is, in the Luce-Hick model the choice probabilities follow the
Luce model of stochastic choice and a monotone transformation of the
decision times (of equiprobable decisions) follows Hick's law.
Compared with the empirical results of \textcite{hick_rate_1952}, the
above definition makes the stronger claim that also the decision times
of non-equiprobable decision processes can be represented by an
entropy but neither requires $T$ to be linear nor the entropy to be in
Shannon form\footnote{Interestingly,
  \textcite['s][]{pieron_sensations_1952} law suggests reaction times in
  binary decisions to depend in power form on the stimulus intensity.
  In many experiments testing this hypothesis
  \parencite{donkin_pieron_2014}, the stimulus is provided in the form
  of a mixing weight (e.g., the proportion of black and white dots as
  in \textcite{ratcliff_modeling_1998}).}, i.e., $r=1$.

In order to characterize the Luce-Hick model via the procedural mixture set
theorem, we require a sufficiently rich set of outcomes to generate decision
processes with arbitrary choice probabilities.
\begin{definition}[Richness of Outcomes]
  For every $x \in \mathscr{X}$ and every $\mu \in [0,1]$ there exists a
  countably infinite number of alternatives $\{y_1,y_2,\ldots\}$ such
  that $p_{\{x,y_i\}}(x)=\mu$.
\end{definition}

We now introduce conditions that (given a rich set of alternatives) are
necessary and sufficient to characterize the Luce-Hick model.
\begin{definition}[Positivity]
  For all $x,y \in \mathscr{X}$, $p_{\{x,y\}}(x)>0$ and $\tau(\{x,y\}) > \tau(\{x\})$.
\end{definition}
Thus, every element of an opportunity set has a nonzero probability of being
chosen and there is a positive deliberation time for the choice
between at least two
items.
\begin{definition}[IIA]
  A stochastic choice function $p$ fulfills {\em IIA at} $x,y \in \mathscr{X}$ if for
  all $C \in \mathscr{C}$ we have that
  \begin{align}
    \label{eq:pathIndependence}
    \frac{p_{\{x,y\}}(x)}{p_{\{x,y\}}(y)} = \frac{p_{C}(x)}{p_{C}(y)}.
  \end{align}
  The stochastic choice function fulfills {\em IIA} if it fulfills IIA for
  all pairs of alternatives.
\end{definition}
Luce's choice axiom states that relative probabilities are unaffected by the
addition of other options. We next impose that comparative decision times are
unaffected by additional options.
\begin{definition}[Independent Decision Times]
  A stochastic choice function $p$ and decision time $\tau$ fulfill independence of
  decision times if for all $C,D,E \in \mathscr{C}$ such that $(C
  \cup D) \cap E = \emptyset$ and $p_{C \cup E}(C) = p_{D \cup E}(D)$ it holds that
  \begin{align}
    \tau(C) \geq & \tau(D) \nonumber\\
    \Leftrightarrow \quad
    \tau(C \cup E) \geq & \tau(D \cup E).
    \label{eq:independentDecisionTimes}
  \end{align}
\end{definition}
This states that the decision time of a decision process is monotone in the
decision time of its subprocesses, i.e., the time it would take to make a choice
from a subset of the alternatives.
\begin{definition}[Continuity of Decision Times]
  For all sequences of sets $\{A^k\equiv\{a_1^k,\ldots,a_n^k\}\}_{k}$ and
  $A = \{a_1,\ldots,a_m\}$, if $p_{A^k}(a_i^k) \rightarrow p_A(a_i)$ for all $i \in \{1,\ldots,m\}$ and
  $p_{A^k}(a_i^k) \rightarrow 0$ for all $i \in \{m+1,\ldots,n\}$ then $\tau(A^k) \rightarrow \tau(A)$.
\end{definition}

\begin{definition}[Sufficiency of Choice Probabilities]

\end{definition}
Continuity has two main implications. First, it imposes that decision
times are continuous in the choice probabilities of alternatives and
only these choice probabilities matter for decision times. Second, it
imposes that given a limit $n$ on the number of alternatives, as
choice probabilities of some ($m-n$ many) alternatives converge to
zero, the mere presence of these alternatives does not affect the
decision times.

The following corollary is now obvious:
\begin{corollary}[Characterization of Luce-Hick Model]
  \label{coro:luceHick}
  Suppose $\mathscr{X}$ fulfills richness in outcomes. Then the
  following statements are equivalent.
  \begin{enumerate}
  \item
    $p$ and $\tau$ fulfill
    positivity,
    IIA,
    independence of decision times, and
    continuity of decision times.
  \item $p$ and $\tau$ form a Luce-Hick model.
  \end{enumerate}
\end{corollary}

Interestingly, for very different reasons Luce was aware of the importance of
his IIA axiom for the use of entropy in psychophysics, writing ``[...]
information theory implicitly presupposes the consequences of [IIA], which
are relatively strong---specifically, when discrimination is imperfect, it means
that choice behavior can be scaled by a ratio scale''
\textcite[][p.12]{luce_individual_1959}.

\section{Comparative Statics}\label{sec:comparativeStatics}
In the value of a procedural mixture, $U(\mu a \oplus (1-\mu) b)$, the parameter $q$ sets
a threshold for $U(a)$ and $U(b)$ that determines whether mixing increases $U$
or not. To make this precise, we introduce a positive and a negative value of
mixing.
\begin{definition}[Value of Mixing]
  $\succsim$ exhibits a {\em negative (positive) value of mixing} at $a \in \mathscr{M}$ if $a \succ (\prec) \mu a
  \oplus (1-\mu) a$.
\end{definition}
The following result is then straightforward:
\begin{proposition}[Monotone Mixing]\label{prop:monotoneMixing}
  If $\succsim$ has a mixture entropy representation, then the
  following statements are equivalent:
  \begin{enumerate}
  \item $\succsim$ exhibits a negative (positive) value of mixing at $a$,
  \item $\succsim$ exhibits a negative (positive) value of mixing at $\mu a \oplus (1-\mu) a$,
  \item $U(a)(r-1) > (<) q$.
  \end{enumerate}
\end{proposition}
From this result follows that if $r>1$, then iteratively
mixing an element with itself yields a sequence of elements for which $U$
converges to $q/(r-1)$. If $r \leq 1$, then $U$ diverges to $\infty$ or $-\infty$, depending
whether for the initial element $U(a)(r-1) \gtrless q$.
\begin{example}
  In our example data of Figure \ref{fig:decisionData}, the value of
  mixing is positive if $T$ is an increasing function. If $r>1$, then
  the mixture entropy $U$ converges to $q$ as the number of options
  increases. If $r \leq 1$, then additional options let $U$ diverge to
  $\infty$. It is noteworthy that this does not mean that decision times
  diverge. Since the representation is ordinal, the actual decision
  times $T \circ U$ may still be bounded in case $\lim_{u\rightarrow \infty} T(u) < \infty$.
  Thus, the limit behavior of decision times alone does not allow us to
  distinguish between $r<1$ and $r>1$.
\end{example}

In addition to setting a threshold for a positive value of mixing, $q$
controls in a procedural mixture (e.g., $U(\mu a \oplus (1-\mu) b)$) the
relative importance of value derived from the mixture weight
($H_r(\alpha)$) compared with the value from the mixed elements ($U(a)$ and
$U(b)$). The relevant comparative statics results are relegated to
Appendix \ref{sec:furtherComparativeStatics} because these results are
only relevant if there are outcomes (i.e., elements that are not
generated from mixtures themselves) $x \succ y$ as the following remark
shows:
\begin{remark}\label{rem:trivialSingletons}
  If $\mathscr{S}$ is generated from finite procedural mixtures of
  members of a set $\mathscr{X}$ and for all $x,y \in \mathscr{X}$,
  $U^1(x)=U^1(y)$, and $U^2(x) = U^2(y)$ then $U^1$ and $U^2$
  represent the same relation if and only if the signs of
  $U^1(x)(r^1-1)- q^1$ and $U^1(x)(r^1-1)- q^1$ are identical and
  $r_1=r_2$.
\end{remark}
It follows from this remark that the magnitude of the parameter $q$
only matters in comparison to a cardinal value of outcomes. If there
do not exist outcomes $x \succ y$, then by the uniqueness properties of
the representation we can find an affine transformation of $U$ such
that the valuation of any existing outcomes is equal to zero. Any
subsequent multiplication of $q$ by a positive factor results in an
increasing linear transformation of $U$ (which does not change the
represented relation).
\begin{example}
  The previous remark is the underlying reason why the Luce-Hick model
  only has a parameter $r$ and no parameter $q$. It is plausible that
  trivial decisions always have the same reaction time and that
  nontrivial decisions take longer than trivial decisions. Thus, if
  $\mathscr{X} \ni x,y$ refers to the set of trivial decisions and $\tau(x)
  = \tau(y)$ for all its elements, then the only meaningful parameter is
  $r$.
\end{example}

We now turn to the interpretation of the parameter $r$. The parameter $r$
controls the degree of the effect of mixing on the value. That is, it controls
how much the value increases (or decreases) by an additional mixing stage.
\begin{definition}[Comparative Value of Mixing]
  $\succsim_1$ yields a {\em higher value of mixing} than $\succsim_2$
  if for all $\alpha, \beta, \gamma < 1/2$ and some $d \in \mathscr{S}$ at which
  $\succsim_1$ and $\succsim_2$ exhibit a positive value of mixing,
  \begin{align}\nonumber
    \alpha d \oplus (1-\alpha) d \quad
    \succsim_1 \quad \beta (\gamma d \oplus (1-\gamma) d ) \oplus (1-\beta) (\gamma d \oplus (1-\gamma) d ),
  \end{align}
  then
  \begin{align}\nonumber
    \alpha d \oplus (1-\alpha) d \quad
    \succsim_2 \quad \beta (\gamma d \oplus (1-\gamma) d ) \oplus (1-\beta) (\gamma d \oplus (1-\gamma) d ).
  \end{align}
\end{definition}
In words, if under $\succsim_2$ a binary mixture has a greater value
than a mixture across four elements, then this must be also the case
under $\succsim_1$.

\begin{proposition}\label{prop:comparativeR}
  Let $\succsim_1$ and $\succsim_2$ be a mixture entropy value with
  representations $U_1$ and $U_2$ and parameters $r_1$, $q_1$ and
  $r_2$, $q_2$, respectively. Then the following statements are
  equivalent.
  \begin{enumerate}
  \item $\succsim_1$ yields a higher value of mixing than $\succsim_2$.
  \item $r_1 \geq r_2$.
  \end{enumerate}
\end{proposition}

\begin{example}
  The decision data of \textcite{hick_rate_1952} suggest that (given
  that $T$ is linear), $r=1$ is a plausible parameter for the decision
  of which one of a number of buttons on a keyboard to press. This
  suggests that the decision times of let's say whether to press a
  button with the left or right hand does not affect the additional
  decision time from choosing whether to press with the index finger
  or the pinkie. In contrast, preferential choices such as the snack
  choices commonly studied in experiments may become increasingly
  complex as the number of alternatives rises. Choices between food
  items may be relatively simple between two items but may become
  increasingly complex as additional options are added. Preferential
  choices would then exhibit a higher value of mixing, leading to a
  different parameter $r$.
\end{example}

\section{Disjoint Independence}\label{sec:disjointIndependence}
The present section shows that a similar representation theorem as
Theorem \ref{thm:ProceduralMixture} holds for mixture sets when
weakening the independence axiom instead of relaxing reducibility.
Intuitively, procedural mixtures treat the mixture components as if
they were distinct even when the mixed components are identical. We
therefore restrict the independence axiom to hold only for distinct
elements in order to obtain an analogous result on mixture sets. The
main upside of this result is that it does not require the ``mixture
richness'' of procedural mixture sets which require infinitely many
distinct elements but this comes at the cost of notational elegance.

\begin{example}
  The Luce model of stochastic choice is restrictive. In some cases,
  the associative structure generated by the Luce model only holds for
  a subset of the decisions. To show this, Figure
  \ref{fig:disjointData} exemplifies the well-known red bus/blue bus
  paradox of \textcite{debreu_review_1960}.

\begin{figure}[ht]
  % These decision times have been obtained by parameters r=1, q=1/ln(2), U(x)=.5
  \centering
  \begin{tabular}{ccc|c}
    Option 1 & Option 2 & Option 3 & Duration \\
    \hline
    {\sc Airplane} & {\sc Blue Bus} & {\sc Car} & \multirow{2}{*}{1.9s}\\
    20\% & 20\%& 60 \% & \\
    {\sc Airplane} & {\sc Blue Bus} & - & \multirow{2}{*}{1.5s}\\
    50\% & 50\% & & \\
    {\sc Blue Bus} & {\sc Red Bus} & {\sc Car} & \multirow{2}{*}{1.4s}\\
    12.5\% & 12.5\%& 75\% & \\
    {\sc Airplane} & {\sc Car} & - & \multirow{2}{*}{1.3s}\\
    25\% &  75\% & & \\
    {\sc Blue Bus} & {\sc Car} & - & \multirow{2}{*}{1.3s}\\
    25\% &  75\% & & \\
    {\sc Red Bus} & {\sc Car} & - & \multirow{2}{*}{1.3s}\\
    25\% &  75\% & & \\
    {\sc Blue Bus} &  {\sc Red Bus} & - & \multirow{2}{*}{1.0s}\\
    50\% & 50\%& & \\
    {\sc Airplane} & - & - & \multirow{2}{*}{0.5s}\\
    100\% & & & \\
    {\sc Red Bus} &  - & - & \multirow{2}{*}{0.5s}\\
    100\% & & & \\
  \end{tabular}
  \caption{Choice Probabilities and Decision Time Data in Nested
    Stochastic Choice}
  \label{fig:disjointData}
\end{figure}

The choice probabilities violate Luce's IIA axiom, because the
relative choice probability of {\sc Blue Bus} and {\sc Car} depends on
the presence of a {\sc Red Bus} among the options: relative to the
{\sc Car}, the choice of a {\sc Blue Bus} becomes less likely after a
{\sc Red Bus} is added to the options.

The decision time relation also violates independence because the
decision between an {\sc Airplane} and a {\sc Blue Bus} takes longer
than a decision between a {\sc Blue Bus} and a {\sc Red Bus}, despite
the singleton decisions for an {\sc Airplane} and for a {\sc Red Bus}
taking equally long.

However, such behavior is reasonable. If the decision maker does not
care about the color of the bus, the decision between the red bus and
the blue bus is trivial and may take less time than the more
meaningful decision between an airplane and a bus. Therefore, in the
upcoming result we maintain that the independence axiom only holds for
sufficiently distinct options. In the above example, both IIA and
independence of decision times still hold among the three options {\sc
  Airplane}, {\sc Blue Bus}, and {\sc Car} but not among the options
{\sc Red Bus}, {\sc Blue Bus}, and {\sc Car}.
\end{example}

We introduce additional notation: Let $\mathscr{Z}$ be a set. For a
given mixture set $\mathscr{M}$, a {\em support} is a function $supp:
\mathscr{M} \rightarrow \mathscr{Z}$ that fulfills for all $a,b \in
\mathscr{M}$ and all $\alpha \in (0,1)$: $supp(\alpha a \oplus (1-\alpha) b) = supp(a) \cup
supp(b)$. $\langle \mathscr{M}, supp \rangle$ is a {\em supported mixture set} if
the image $supp(\mathscr{M})$ is closed under nonempty intersections
and under nonempty relative complements. A subset $Z$ of $\mathscr{Z}$
is {\em essential} if there exist $a,b \in \mathscr{M}$ such that
$supp(a) \subseteq supp(b) = Z$ and $a \not\sim b$.

\begin{example}
  In the example, we may consider {\sc Blue Bus} and {\sc Car} to have
  disjoint support but {\sc Blue Bus} and {\sc Red Bus} to not have
  disjoint support. This allows us to specify for which elements of
  the mixture set independence applies and for which it does not.
\end{example}

\begin{axiom}[Disjoint Independence]\label{axiom:disjointIndependence}
  For all $a,a',b\in \mathscr{M}$, $\mu
  \in (0,1)$, if $(supp(a) \cup supp(a')) \cap supp(b) = \emptyset$ then
  \begin{align}
    a \succsim & a' \nonumber\\
    \Leftrightarrow \quad \mu a \oplus (1-\mu) b \succsim & \mu a' \oplus (1-\mu) b.
  \end{align}
\end{axiom}

We can embed a mixture set $\mathscr{M}$ with a relation $\succsim^*$
fulfilling Axioms 1,2, and \ref{axiom:disjointIndependence} partly
into a procedural mixture set $\mathscr{S}$ with a relation $\succsim$
fulfilling Axioms 1-3. If there are sufficiently many essential
subsets in the support, the embedded part has the same uniqueness
properties as the mixture entropy representation and we obtain the
following theorem:
\begin{theorem}[Procedural Mixture Set Embedding]
  \label{thm:disjointMixture}
  Let $\langle \mathscr{M}, supp \rangle$ be a supported mixture set and
  $\succsim^*$ be a binary relation on $\mathscr{M}$ such that there
  exist at least three disjoint and essential subsets of
  $\mathscr{X}$.

  Then the following statements are equivalent:
  \begin{enumerate}
  \item The relation $\succsim^*$ fulfills Axioms \ref{axiom:weakOrder},
    \ref{axiom:continuity}, and \ref{axiom:disjointIndependence}.
  \item There exist parameters $q\in \mathbb{R},r\in \mathbb{R}_{++}$, and a function
    $U:\mathscr{M}\rightarrow \mathbb{R}$ representing $\succsim^*$ such that
  \begin{align}
    U(\mu a \oplus (1-\mu)b) =& \mu^{r} U(a) + (1-\mu)^r U(b) + q \cdot H_r(\mu)\label{eq:disjointEntropyRepresentation}\\
    H_r(\mu) =&
              \begin{cases}
                -\mu \ln \mu - (1-\mu)\ln (1-\mu), & r=1\\
                \frac{1-\mu^r  -(1-\mu)^r}{r-1}, & r\neq 1
              \end{cases}
  \end{align}
  if $supp(a) \cap supp(b) = \emptyset$. $U$ is continuous in mixture weights.
  \end{enumerate}
\end{theorem}
\textcite{chen_measuring_2020} prove a similar result for four
disjoint subsets using a different proof technique.

we now combine Theorem \ref{thm:ProceduralMixture} and Theorem
\ref{thm:disjointMixture} to characterize a model of decision times
for nested stochastic choice \parencite{kovac_behavioral_2022}.
Concretely, we will characterize the following model:

\begin{definition}[Nested Luce-Hick Model]
  A stochastic choice function $p$ and a decision time $\tau$ form a
  {\em nested Luce-Hick model} if there exists
  a continuous function $v: \mathscr{C} \rightarrow \mathbb{R}_{+}$ with $v(C)=0$ iff
  $C=\emptyset$,
  a unique parameter $r \in \mathbb{R}_{++}$,
  a unique strictly monotone continuous function $T: \tau(\mathscr{C}) \rightarrow \mathbb{R}$,
  a unique partition $\mathscr{S}$ of $\mathscr{X}$,
  for all $S \in \mathscr{S}$
  a unique set of parameters $r_S \in \mathbb{R}_{++}$,
  and strictly monotone continuous functions
  $t_S: \mathbb{R}_{+} \rightarrow \mathbb{R}$ with $t_S(0)=0$ which are unique
  up to joint linear transformations,
  such that for all sets $D \in \mathscr{C}$:
    \begin{align}
      \label{eq:nestedluceModel}
      p_D(x) =
      \begin{cases}
        \frac{v(D \cap S)}{\sum_{S' \in \mathscr{S}} v(D \cap S')}
        \frac{v_S(x)}{\sum_{y \in S} v_S(y)}, & x \in D\\
        0, & \text{else,}
      \end{cases}
    \end{align}
    \begin{align}
      \label{eq:nestedhickModel}
      T(\tau(D))
      = & \sum_{S \in \mathscr{S}: S \cap D \neq \emptyset}  p_D(S)^{r} \cdot t_S\left(H_{r_S}\left(p_{D \cap S}(x)\right)\right) \nonumber\\
          & + H_r\left(\left(p_D(S)\right)_{S \in \mathscr{S}:p_D(S)>0}\right).
    \end{align}
\end{definition}
Thus, for each category the decision maker has an entropy
representation of the decision time contributions of choice within the
category and the choice between categories also takes time according
to an entropy representation. This model is highly flexible. It can
capture that decision times across very different alternatives may
take longer than choices between very similar alternatives or the
reverse. It can also capture that within some but not necessarily all
categories, choice becomes increasingly complex the more options are
added or that choice within some category is much faster than in
others.
\begin{example}
  In Figure \ref{fig:disjointData}, the decision times are generated
  by a partition $\mathscr{S} = \{\{\textsc{Airplane}\},\{\textsc{Blue
    Bus}, \textsc{Red Bus}\},\{\textsc{Car}\}\}$ $r=r_S=1$,
  $T(t)=0.5+t/\ln 2$, $t_S(t)=t/2$ for all $S \in \mathscr{S}$.
  Effectively, the model treats the decision between categories the
  same way as the model in the introductory example. The decision
  within each category is however much faster, by a factor of 2.
\end{example}

As \textcite{kovac_behavioral_2022} show, the categories can be
identified from whether IIA holds for pairs of options:
\begin{definition}[Categorical Similarity]
  Two alternatives $x$ and $y$ are {\em categorically similar} if the
  stochastic choice function satisfies IIA at $x,y$. Alternatives $x$
  and $y$ are {\em equally similar} to $z$ if either both $x$ and $y$
  are categorically similar to $z$ or if neither $x$ nor $y$ are
  categorically similar to $z$.
\end{definition}
Categorical similarity is an equivalence relation and therefore we can
partition $\mathscr{X}$ into sets of categorically similar
alternatives. The nested stochastic choice rule that governs the
decision probabilities fulfills the following axiom of
\textcite{kovac_behavioral_2022}.
\begin{definition}[IIA for Equally Dissimilar Alternatives]
  Whenever $x$ and $y$ are equally similar to $z$ , then for all
  sets $A$ containing $x,y$:
  \begin{align}
    \label{eq:IIASymmetric}
    \frac{p_A(x)}{p_A(y)} = \frac{p_{A \cup \{z\}}(x)}{p_{A \cup \{z\}}(y)}.
  \end{align}
\end{definition}

The following new axiom is the decision time counterpart to the
previous axiom.
\begin{definition}[Decision Time Independence for Equally Dissimilar Alternatives]
  Whenever all elements of $C$ are equally dissimilar to all elements
  in $A \cup B$, and $p_{A \cup C}(A)=p_{B \cup C}(B)$ then:
  \begin{align}
    \tau(A) \geq & \tau(B) \nonumber\\
    \Leftrightarrow \quad
    \tau(A \cup C) \geq & \tau(B \cup C)
    \label{eq:DTIndependenceSymmetric}
  \end{align}
\end{definition}

\begin{definition}[Category-wise Continuity of Decision Times]
  For all sequences of sets $\{A^k\equiv\{a_1^k,\ldots,a_n^k\}\}_{k}$ and
  $A = \{a_1,\ldots,a_m\}$, if for all $S \in \mathscr{S}$
  $p(A^k \cap S) \rightarrow p(A \cap S)$ and
  $p_{A^k \cap S}(a_i^k) \rightarrow p_{A \cap S}(a_i)$ for all $i \in \{1,\ldots,m\}$ and
  $p_{A^k \cap S}(a_i^k) \rightarrow 0$ for all $i \in \{m+1,\ldots,n\}$ then $\tau(A^k) \rightarrow \tau(A)$.
\end{definition}
In other words, if the choice probabilities of all categories converge
and the choice probabilities within each category converge, then the
decision time converges as well.

To generate a procedural mixture set within each category, we assume
the following richness condition.
\begin{definition}[Richness]
  For each $x \in \mathscr{X}$ and every $\alpha \in (0,1)$ there exists a
  countably infinite number of categorically similar alternatives $y$
  and a categorically dissimilar alternative $z$ such that
  $p_{\{x,y\}}(x) = \alpha = p_{\{x,z\}}(x) $.
\end{definition}

Thus, in the nested logit we have an associative structure not only in
the disjoint attribute case but also in the identical attribute case
and can use Theorem \ref{thm:ProceduralMixture} to embed a procedural
mixture set to characterize decision times which may have different
parameters $r$ and $q$ for different subsets of alternatives.

\begin{corollary}[Characterization of Nested Luce-Hick
  Model]\label{coro:nestedLuceHick}
  Suppose $p,\tau$ has at least 3 equivalence classes of categorically
  similar alternatives and fulfills richness. Then the
  following statements are equivalent:
  \begin{enumerate}
  \item $p,\tau$ is a nested Luce-Hick model.
  \item $p,\tau$ fulfill
    positivity,
    category-wise continuity of decision times,
    IIA for equally dissimilar alternatives, and
    decision time independence for equally dissimilar alternatives.
  \end{enumerate}
\end{corollary}

\section{Literature}\label{sec:literature}

There are three branches of literature related to the present paper; a
literature on axiomatic characterizations of entropies, a literature on axiomatic
characterizations of decision times, and a literature on mixture sets and
relaxations of the reduction of compound mixtures axiom.

\subsection{Characterizations of Entropy Functions}
For a general survey of the literature of the characterization of information
measures, see \textcite{csiszar_axiomatic_2008}.

\textcite[ch. 3.12]{krantz_foundations_1971} defined entropy structures and
showed that a relation represented by $H_r$ fulfills the assumptions of an
entropy structure. However, they did not provide a characterization result of
$H_r$ or $H_1$. Their operation $\circ$ of an entropy structure captures the idea of
$a \circ b$ denoting the physical system consisting of two independent physical
systems $a$ and $b$. The mixture operation $\mu a \oplus (1-\mu) b$ instead better
applies to the mixture of distinguishable gases or liquids $a$ and $b$ with
proportion $\mu$ and thus our representation captures the so-called {\em entropy of
  mixing}.

Closely related to the present paper is \textcite{luce_utility_2008} in which
the utility of gambling is characterized as expected utility plus the entropy of
the lottery. There are three technical improvements the present paper makes.
First, \textcite{luce_utility_2008} assume the existence of a status quo
consequence and directly impose that the utility of a gamble between the status
quo and some outcome ocurring in some event is separable. Second,
\textcite{luce_utility_2008} assume that outcomes and gambles are closed under
an operation they call ``joint receipt'', interpreted as receiving two gambles
simultaneously. They further assume that the utility over the two received
gambles is additive, i.e., that the utility of the joint receipt of lotteries is
the sum of the individual lotteries. Preferences over the gambles are therefore
independent and thus the decision maker's risk attitude over one gamble may not
be influenced by whether the second gamble is risky or not. Third,
\textcite{luce_utility_2008} assume the existence of kernel equivalents. The
kernel equivalent of a gamble is an outcome that when received simultaneously
with an event-resolving but payoff-irrelevant gamble leaves the decision maker
indifferent. Overall, their axioms are somewhat nonstandard and lack the
accessibility of the mixture sets introduced in
\textcite{herstein_axiomatic_1953}.

We show that only small adjustments to the standard axioms need to be made to
obtain entropy measures as a utility component. We do not assume the joint
receipt of gambles or the existence of a status quo outcome. Additive
separability instead naturally arises from the von Neumann-Morgenstern
independence axiom. While \textcite{luce_utility_2008} assumes that certainty
equivalents and kernel equivalents exist, our model and axiomatization are
consistent with the nonexistence of certainty equivalents such as in the case
when the mixture set is generated starting from mixtures of a finite set of
alternatives, mixtures of these mixtures, etc..

The literature on rational inattention has provided characterizations of
expected utility with entropy costs of attention
\parencite{caplin_rationally_2017,de_oliveira_rationally_2017}.
\textcite{ellis_foundations_2018,lin_stochastic_2020,lu_random_2016}
characterize more general information cost functions. Most of this literature
relies on observations of choices over menus or alternatives and treats choices
over information structures as unknown.

\subsection{Decision Processes and Axiomatizations of Response Times}
With respect to our application to decision processes, there exists a
large literature on decision processes \parencite{usher_dynamics_2013,
  bogacz_physics_2006} especially the drift-diffusion model
\parencite{ratcliff_theory_1978,ratcliff_diffusion_2016}. Several
studies have proposed models for decision times in decision processes
with multiple alternatives
\parencite{baldassi_behavioral_2020,krajbich_multialternative_2011,mcmillen_dynamics_2005,tajima_optimal_2019}.
Another interesting line of research are foundations for decision
times when agents follow particular behavioral decision models. Most
notably, this has been done for dual-process decision making
\parencite{achtziger_fast_2014,alos-ferrer_dual-process_2016},
cognitive sophistication \parencite{alos-ferrer_cognitive_2025},
rational inattention \parencite{fudenberg_speed_2018}, hyperbolic
discounting \parencite{chabris_allocation_2009}, and directed
cognition \parencite{gabaix_costly_2006}. Most closely related to the present paper are the
axiomatic studies discussed in more detail below.

Like the present paper, \textcite{baldassi_behavioral_2020} also
employ the Luce model to characterize decision times. They obtain the
decision times of the drift diffusion model for binary choice by
imposing that across decisions the accuracy (in our notation $\max(\mu,
1-\mu)-1/2$) is proportional to the product of the average decision time
multiplied by the ease of comparison $\ln (\max(\mu, 1-\mu)) - \ln \min(\mu,
1-\mu)$. This {ad hoc} functional imposition generates decision times as
in the drift-diffusion model for binary choices. For multi-alternative
choice, \textcite{baldassi_behavioral_2020} show that the Metropolis-DDM
model \parencite{cerreia_multinomial_2022}, i.e., binary
drift-diffusion comparisons together with Markovian exploration where
choices are made at a fixed deadline, generates softmaximizing behavior.

\textcite{koida_multiattribute_2017} axiomatizes multiattribute
decision times based on time-indexed preferences over lotteries of
goods. Preferences at any point may be incomplete but become more and
more decisive as time progresses and the decision maker resolves
internal conflict over which attributes they care about. Decision
times are interpreted as the moment at which the preference becomes
decisive over two alternatives. Choices and their timing in the model
are deterministic but predictions therefore naturally require the
identification of a large number of parameters.

\textcite{fudenberg_testing_2020} characterize the binary
drift-diffusion model as a joint probability distribution over
decision times and the choice made from two options by imposing that
the stopping time follows that of a Brownian motion and that the
revealed boundary (as a function of the decision probability at any
particular decision time, the average choice imbalance across time,
and the average decision time) is nonnegative for all decision times.

\textcite{echenique_response_2017} axiomatize response times for
binary choice data. Similar to the present study, they obtain a
representation of response times instead of a direct characterization
of the functional form of response times. Different from the present
study, they work with deterministic choices and distinguish between
the response time to choose $a$ over $b$ and the response time to
choose $b$ over $a$. Most importantly, they address the issue of
finite data while the present study requires a rich data set
fulfilling (at least for a subset of the options) the IIA axiom.

\subsection{Mixture Sets and the Reduction of Compound Mixtures Axiom}
There is a vast literature of decision theoretic papers that employ the mixture
sets introduced by \textcite{herstein_axiomatic_1953}. Commonly the axioms on
preferences are being varied instead of the structure of the mixture set. In
contrast, \textcite{mongin_note_2001} examines under which conditions mixture
sets can be treated as convex subsets of a vector space.

The reduction of compound mixtures axiom has received substantial attention.
The literature on recursive utility models following
\textcite{kreps_temporal_1978} analyzes intertemporal decision problems in which
within each time period the reduction of compound mixtures axiom holds but
between time periods it does not. A large literature following
\textcite{segal_two-stage_1990} remove the reduction of compound mixtures
assumption completely and study two-period mixtures under various axioms on
preferences. In contrast to their work, this paper studies mixtures with an
arbitrary, finite number of stages and maintains that the order of resolution of
compounding is irrelevant while the mixing itself is not.

\section{Conclusion}\label{sec:discussion}
The present analysis provides a foundation for the study of violations
of betweenness due to procedural aspects. The entropy representation
is obtained by relaxing the assumption that mixtures of an element
with itself yields the same element or by weakening the independence
axiom to mixtures of sufficiently distinct elements. Entropy measures
play an important role in a large number of applications and the
simple axiomatization provided in this paper may prove useful in other
contexts.\footnote{For example, in the
  context of decisions under risk, uncertainty effects may induce violations of
  betweenness \parencite{gneezy_uncertainty_2006}. This suggests that
  individuals may also attach a ``procedural'' value to the uncertainty of
  lotteries.}

The application to decision processes provides a parsimonious model of the
relation between choice probabilities and decision times. It is perhaps striking
that the central prediction of the drift-diffusion model --- that decision times
are monotone in how even the choice probabilities are --- is obtained from very
simple assumptions about decision times for choices between multiple
alternatives. However, the result can also be understood as an impossibility
result; if one accepts that the decision time of a decision process should
increase in the decision times of its subsets and the choice probabilities
fulfill IIA, then one has to accept that decision times are related to choice
probabilities in the somewhat restricted functional form of an entropy. While
the representation is ordinally consistent with the binary drift-diffusion
model, the very different functional form highlights the difficulty of extending
the drift-diffusion model to multiple options.

By the example of nested stochastic choice we have shown that
procedural mixtures and/or disjoint independence provides a starting
point for characterizations in which the entropy represents decision
times only on the domain of choices where the Luce model is plausible.
There is much recent interest in the literature on axiomatic
foundations of variations of the Luce model, e.g., the mixed logit
\parencite{saito_axiomatization_2018}, the nested logit
\parencite{kovac_behavioral_2022}, or the conditional logit
\parencite{breitmoser_axiomatic_2021}. Axiomatically studying decision
times for such models would be an interesting avenue for future
research given the prevalence of stochastic choice in empirical
applications.

\section*{Acknowledgments}
This work was financially supported by the Center for Research in
Econometric Theory and Applications (Grant No. 109-L900-203) from the
Featured Areas Research Center Program within the framework of the
Higher Education Sprout Project by the Ministry of Education (MOE) in
Taiwan, and by the Ministry of Science and Technology (MOST), Taiwan,
under Grant No. 107-2410-H-002-031 and 109-2634-F-002-045.
\appendix
\section{Proof of Theorem \ref{thm:ProceduralMixture}} \label{sec:proof}
\begin{proof}
Neccessity is straightforward. We prove sufficiency.

Let $\mathscr{Q}=\mathscr{S}/\sim$ be the quotient set of $\mathscr{S}$ with respect to the equivalence relation
$\sim$. Note that whenever $a,b \in \mathscr{S}$ and $a \sim b$, then any utility representation
$U$ must fulfill $U(a) = U(b)$. Note further that $\mathscr{Q}$ is a procedural mixture
set when endowed with $\succsim^*$ such that $q \succsim^* r$ if and only if $a \succsim b$ for some
$a \in q$ and $b \in r$. Thus, finding a utility on $\mathscr{Q}$ is equivalent to finding a
utility on $\mathscr{S}$. We therefore assume for the remainder of the proof that $\mathscr{S}=\mathscr{Q}$.

Let the order topology on $\mathscr{S}$ be the topology generated by the
subbase of upper and lower contour sets of the asymmetric part of $\succsim$.
\begin{lemma}
$\mathscr{S}$ is topologically connected under the order topology.
\end{lemma}
\begin{proof}
  If $\mathscr{S}$ is not connected, then it is the union of two nonempty disjoint open
  sets $\mathscr{S}'$ and $\mathscr{S}''$. Take any elements $s'\in \mathscr{S}'$ and $s''\in \mathscr{S}''$. The set
  $\mathscr{S}'''=\{a|\exists \mu: a=\mu s' \oplus (1-\mu)s''\}$ is disconnected in the subspace topology
  by the disjoint nonempty open sets $S'\cap S'''$ and $S''\cap S'''$. Since the upper
  and lower contour sets of $\succsim$ form a subbase of $\mathscr{S}$, the upper and lower
  contour sets of $\succsim$ in $\mathscr{S}'''$ form a subbase of the subspace topology. By
  Axiom 2, the bijection $f:\mu \mapsto \mu s' \oplus (1-\mu)s''$ is then continuous. But then
  the preimages $f^{-1}(\mathscr{S}'\cap \mathscr{S}'')$ and $f^{-1}(\mathscr{S}''\cap \mathscr{S}''')$ are open, disjoint,
  and disconnect the unit interval, a contradiction.
\end{proof}

\begin{lemma}\label{lemm:Coseparability} $\succsim$ is coseparable, i.e.,
\begin{align}
    \mu a \oplus (1-\mu) b &\sim \bar \mu \bar a \oplus (1-\bar \mu) \bar b\\
    \mu a' \oplus (1-\mu) b &\sim \bar \mu \bar a' \oplus (1-\bar \mu) \bar b\\
    \mu a \oplus (1-\mu) b' &\sim \bar \mu \bar a \oplus (1-\bar \mu) \bar b'
\end{align}
jointly imply
\begin{align}
    \mu a' \oplus (1-\mu) b' &\sim \bar \mu \bar a' \oplus (1-\bar \mu) \bar b'
\end{align}
\end{lemma}
\begin{proof}
Using commutativity and associativity it is straightforward to show that
\begin{align}\label{eq:cosepmix}
    &1/2 [\mu a \oplus (1-\mu) b ] \oplus 1/2 [\mu a' \oplus (1-\mu) b']
    \\
    =&1/2 [\mu a' \oplus (1-\mu) b ] \oplus 1/2 [\mu a \oplus (1-\mu) b']
\end{align}
for any $\mu, a, b, a',b'$.
Using Axiom 3 together with the assumptions stated above then guarantee the desired result.
\end{proof}
\begin{lemma}
$\succsim$ can be represented by continuous $U$, $F$ such that
\begin{align}
    U(\mu a \oplus (1-\mu) b) = F(a, \mu) + F(b, 1-\mu)
\end{align}
\end{lemma}
\begin{proof}
  We either obtain the representation trivially, if $a \sim b$ for all $a,b \in \mathscr{S}$ or
  using the main theorem of \textcite{qin_quasi-separable_2018} which provides a
  representation theorem for weak orders on (open subsets of) $\mathscr{X}\times \mathscr{Y}\times \mathscr{Z}$ with the
  representation $f(x,z) + g(y,z)$. Here we choose $\mathscr{X}=\mathscr{S}$, $\mathscr{Y}=\mathscr{S}$, and $\mathscr{Z}=(0,1)$
  and endow the space with the product topology of the order topologies and the
  subspace topology of the reals. Thus, we will first obtain the representation
  on $\mu \in (0,1)$ and then extend it to $[0,1]$ using Axiom 2. To apply the main
  theorem of \textcite{qin_quasi-separable_2018}, we require the following
  conditions: essentiality, conditional independence of the $\mathscr{X}$ and $\mathscr{Y}$
  dimension given $\mathscr{Z}$, coseparability of $\mathscr{X}$ and $\mathscr{Y}$ given $\mathscr{Z}$, continuity in
  the product topology, and topological connectedness of $\mathscr{X}$, $\mathscr{Y}$, and $\mathscr{Z}$.

  Since we have a product space, the well-behavedness assumptions of
  \textcite{qin_quasi-separable_2018} are not needed and we also only need
  essentiality instead of strict essentiality. Essentiality requires that for at
  least some $\mu$ and some $a$, then there exist some $b,b'$ such that $\mu a \oplus
  (1-\mu) b\not \sim \mu a \oplus (1-\mu) b'$ and for some $a,b$ there exist some $\mu,\mu'$ such
  that $\mu a \oplus (1-\mu) b\not \sim \mu' a \oplus (1-\mu') b$. The former is guaranteed by Axiom
  3 and the exclusion of the case $a\sim b$ for all $a,b\in \mathscr{S}$. The latter is
  guaranteed by Axiom 2 and the exclusion of the case $a\sim b$ for all $a,b\in \mathscr{S}$.
  Next, we need conditional independence of the $\mathscr{X}$ and $\mathscr{Y}$ dimensions
  for fixed $\mathscr{Z}$ dimension. This holds by Axiom 3. Further,
  coseparability of the $\mathscr{X}$ and $\mathscr{Y}$ dimension given $\mathscr{Z}$ has been shown above.
  Continuity of $\succsim$ holds in the order topology on $\mathscr{S}$. However, we require
  continuity in the product topology on $\mathscr{S}\times \mathscr{S}\times (0,1)$. By Axioms 2 and 3 the
  product topology is finer than the order topology on $\mathscr{S}$, guaranteeing
  continuity in the product topology. Topological connectedness of the product
  topology follows from the connectedness of its components $\mathscr{X}$, $\mathscr{Y}$, and $\mathscr{Z}$.
  The interval $(0,1)$ is obviously connected and each component $\mathscr{S}$ is
  connected in the order topology.

  From \textcite{qin_quasi-separable_2018} then follows the existence of
  functions $F$ and $E$ such that $\succsim$ can be represented by
  \begin{align}
    U(\mu a \oplus (1-\mu) b) = F(a, \mu) + E(b, \mu)
  \end{align}
  Commutativity of the mixture set guarantees that we can redefine $E$ and $F$
  such that $E(b,\mu) = F(b,1-\mu)$. To see this, note that from $U(\mu a \oplus (1-\mu) b) =
  U((1-\mu) b \oplus \mu a)$ follows that $F(b, 1-\mu)+E(a, 1-\mu) = F(a, \mu) + E(b, \mu)$.
  Since this holds for all $a$, $E(b, \mu) = F(b, 1-\mu) + E(a^*, 1-\mu) - F(a^*, \mu)$
  for some arbitrarily chosen $a^*$. Thus, $E(b, \mu) = F(b, 1-\mu) + f(\mu)$ where
  $f(\mu) \equiv E(a^*, 1-\mu) - F(a^*, \mu)$. Substituting in the original representation,
  $U(\mu a \oplus (1-\mu) b) = F(a, \mu) + F(b, 1-\mu) + f(\mu)$. Redefining $\hat F(a, \mu) =
  F(a, \mu) + \begin{cases}f(\mu)/2 & \mu < 1/2 \\ f(1-\mu)/2 & \mu \geq 1/2. \end{cases}$,
  it follows that $U(\mu a \oplus (1-\mu) b) = \hat F(a, \mu) + \hat F(b, 1-\mu)$.
\end{proof}

\begin{lemma}
$F(a,\mu) = A(\mu)U(a) + B(\mu)$ for all $\mu$ and all $a\in \mathscr{S}$.
\end{lemma}
\begin{proof}
For fixed $\mu$, $F(a,\mu)$ is a monotone transformation of $U$:
\begin{align}
    & F(a,\mu) \geq F(b,\mu) \\
    \Leftrightarrow \quad& F(a,\mu) + F(c,1-\mu) \geq F(b,\mu) + F(c,1-\mu)\\
    \Leftrightarrow \quad& \mu a \oplus (1-\mu) c \succsim \mu b \oplus (1-\mu) c \\
    \Leftrightarrow \quad& a\succsim b \\
    \Leftrightarrow \quad& U(a) \geq U(b)
\end{align}
Therefore, we can write $F(a,\mu) = G(U(a),\mu)$. We obtain from associativity:
\begin{align}
  &U(\mu a \oplus (1-\mu) [\lambda b \oplus (1-\lambda) c])\\
  = &G(U(a),\mu) + G(G(U(b),\lambda) + G(U(c),1-\lambda),1-\mu)
      \label{eq:AffinityFromAssociativity}\\
  = &G(U(b),(1-\mu) \lambda)
      + G(G(U(a),\frac{\mu}{1-(1-\mu)\lambda})
      + G(U(c),\frac{(1-\mu)(1-\lambda)}{1-(1-\mu)\lambda}),1-(1-\mu)\lambda)\\
  = &U((1-\mu) \lambda b \oplus (1-(1-\mu)\lambda)
      [\frac{\mu}{1-(1-\mu)\lambda} a \oplus (1-\lambda)\frac{(1-\mu)(1-\lambda)}{1-(1-\mu)\lambda} c])
\end{align}
Noting that we have two continuous additive representations over $\mathscr{S}\times \mathscr{S}$
(specifically here the elements $a$ and $b$), by the uniqueness of additive
representations, we have that $G(\cdot, 1-\mu)$ in
\eqref{eq:AffinityFromAssociativity} is positively affine in its first argument.
Since $b,c$ and $\mu,\lambda$ are arbitrary, this holds for all utility levels.
Therefore $G(U(a),\mu) = A(\mu) U(a) + B(\mu)$ for all $a,b \in \mathscr{S}$ and $\mu \in [0,1]$.
\end{proof}

\begin{lemma}$A(\mu) = \mu^r$, $r\in \mathbb{R}_{++}$.
\end{lemma}
\begin{proof}
We define $H(\mu) = H(1-\mu) = B(\mu) + B(1-\mu)$. Using associativity, we can derive that
\begin{align}
  & A(\lambda)\left[A(\mu) U(a) + A(1-\mu)U(b) + H(\mu)\right] + A(1-\lambda)U(c) + H(\lambda)\\
  = &A(\lambda \mu) U(a) + H(\lambda\mu) \nonumber\\
  &+ A(1-\lambda \mu) \left[A\left(\frac{\lambda (1-\mu)}{1-\lambda \mu}\right)U(b) + A\left(\frac{1-\lambda}{1-\lambda\mu}\right)U(c) + H\left(\frac{\lambda (1-\mu)}{1-\lambda \mu}\right)\right]
\end{align}
Consider a substitution $a'$ for $a$ under which the above condition needs to
still hold. If $\Delta U = U(a)-U(a')$, then it follows that
\begin{align}
    A(\lambda)A(\mu) \Delta U = A(\lambda \mu) \Delta U
\end{align}
and therefore $A$ is multiplicative. Using Cauchy's functional equation it is
straightforward to derive that $A(\mu)=\mu^{r}$, $r \in \mathbb{R}$. By Axiom 3,
$r>0$.
\end{proof}
We now finish the proof. We obtain
\begin{align}
    & \lambda^r H(\mu) + H(\lambda)
    =  (1-\lambda \mu)^r \left[ H\left(\frac{\lambda (1-\mu)}{1-\lambda \mu}\right)\right] + H(\lambda \mu)
\end{align}
and substitute: $\lambda = 1-x$ and $\lambda\mu = y$. Using $H(x)=H(1-x)$ we obtain:
\begin{align}
    & (1-x)^r H\left(\frac{y}{1-x}\right) + H(x)
    =  (1-y)^r H\left(\frac{x}{1-y}\right) + H(y)
\end{align}
with two types of solutions \parencite{ebanks_generalized_1987}:
\begin{align}
    A(\mu) = \mu ;&\quad H(\mu) = -(\mu\ln \mu + (1-\mu)\ln (1-\mu))q + s\\
    A(\mu) = \mu^r;&\quad H(\mu) = -(\mu^r + (1-\mu)^r -1)q + s
\end{align}
where $q,s\in \mathbb{R}$. From Axiom 2 and connectedness, we also have that in
both representations $s=0$. We have therefore obtained the desired
representation:
\begin{align}
    U(\mu a \oplus (1-\mu)b) =& \mu^r U(a) + (1-\mu)^r U(b) + q \cdot H_r(\mu)\\
    \text{ with } H_r(\mu) =&
    \begin{cases}
    -\mu\ln \mu - (1-\mu)\ln (1-\mu), & r=1\\
    -\mu^{r}  -(1-\mu)^{r}+1, & r\neq 1
    \end{cases}
\end{align}
Regarding uniqueness, note that if preferences are nontrivial, then we
immediately have an additively separable preference $U(\mu x \oplus (1-\mu) y)$ over a
continuum of $x$ and $y$ and thus $U$ is unique up to affine transformations.
The uniqueness properties of $r$ and $q$ follow immediately.
\end{proof}

\section{Proof of Corollary \ref{coro:luceHick}}
\begin{proof}
  We prove sufficiency:
  The characterization of the Luce model is standard. We fix some $y \in \mathscr{X}$ and
  define $v(y) = 1$ and $v(x) = \ln{p(x,\{x,y\})} - \ln{1-p(x,\{x,y\})}$. It is
  straightforward to then show that \eqref{eq:luceModel} holds.

  We form an equivalence relation $\approx$ on $\mathscr{C}$ such that $\mathscr{C} \approx \mathscr{D}$ if and only if
  there exists enumerations $C = \{x_1,\ldots,x_n\}$ and $D = \{y_1,\ldots,y_n\}$ such
  that $p(x_i,C) = p(y_i,D)$ for all $i \in \{1,\ldots,n\}$. From Continuity of
  Decision Times follows that if $C \approx D$, then $\tau(C)=\tau(D)$. Each element of
  $\mathscr{C}/\approx$ can be represented by a finite tuple $(\mu_1,\ldots,\mu_n)$ with $\sum_{i}\mu_i =1$,
  $\mu_i \in (0,1]$, and the convention that $\mu_i \geq \mu_{i+1}$ for all $i \in
  \{0,\ldots,n\}$. Notice that this makes a statement such as $C \in (\mu_1,\ldots,\mu_n)$
  meaningful; it means that the set $C$ is an element of the equivalence class
  represented by $(\mu_1,\ldots,\mu_n)$. We endow the set $\mathscr{C}/\approx$ with an operation $\oplus$
  such that $\mu (\mu_1,\ldots,\mu_n) \oplus (1-\mu) (\lambda_1,\ldots,\lambda_k)$ is the tuple obtained from
  rearranging $(\mu \mu_1,\ldots,\mu \mu_n, (1-\mu) \lambda_1, \ldots, (1-\mu) \lambda_k)$ into descending order.
  We further endow the mixture set $\langle\mathscr{C}/\approx, \oplus, =\rangle$ with the weak order induced by
  $\tau$: $a \succsim b$ if there exist $C \in a$ and $D \in b$ such that $\tau(C)\geq \tau(D)$. By
  Continuity of Decision Times, indeed $a \succsim b$, $C \in a$ and $D \in b$ holds if and
  only if $\tau(C) \geq \tau(D)$.

  Notice that if $C \cap E = \emptyset$, $p(C,C \cup E) = \mu$, $C \in a$, and $E \in b$, then
  $C \cup E \in \mu C \oplus (1-\mu) E$. From independence of decision times then follows that
  the relation induced by $\tau$ on the procedural mixture set fulfills
  independence.

  Continuity of $\succsim$ follows straightforward from the fact that $\tau$ is continuous in the
  choice probabilities.

  By Theorem \ref{thm:ProceduralMixture} there exists a representation $U$ of
  $\succsim$ on $\mathscr{C}/\approx$. Since $U$ is continuous and $\tau$ is continuous, there must exist
  a continuous monotone transformation $T$ such that if $C \in b$, then $T \circ \tau(C)
  = U(b)$. Now let $C \cap D = \emptyset$, $C \in a$, $D \in b$ and $\mu = p(C, C \cup D)$. Then $T
  \circ \tau(C \cup D) = U(\mu a \oplus (1-\mu) b) = p(C, C \cup D)^r T \circ \tau (C) + p(D, C \cup D)^r T \circ
  \tau(D) + q H_r(p(C,C \cup D))$. Since $\tau(\{x\}) = \tau(\{y\})$ for all $x,y$ and $U$
  is unique up to affine transformations, we can assume without loss of
  generality that $U((1))=0$, i.e., $T \circ \tau(\{x\}) = 0$. If this is the case,
  then by Positivity and Remark \ref{rem:trivialSingletons} it is without loss
  of generality to assume $q=1$.
\end{proof}

\section{Proof of Proposition \ref{prop:comparativeR}}
For notational convenience, we define $a \equiv \alpha d \oplus (1-\alpha) d$ and $b = \beta c \oplus (1-\beta)
c$, where $c = \gamma d \oplus (1-\gamma)d$.

$a \succ (\prec) d$ if and only if $sgn(q_1-U_1(d)(r_1-1)) = sgn(q_2-U_2(d)(r_2-1)) >
(<) 0$.

By the uniqueness properties of $U$, $a \succsim b$ if and only if
\begin{align}
  \label{eq:ComparativeMixingCondition}
  sgn(q-U(d)(r-1))
  H_r(\alpha)
  \geq
  sgn(q-U(d)(r-1))
  ((\gamma^r + (1-\gamma)^r)H_r(\beta) + H_r(\gamma))
\end{align}
Thus, $a \succsim_1 b$ implies $a \succsim_2 b$ and $a \succ b_1$ implies $a \succ_2 b$ if one of the
following is true: $q_1-U_1(d)(r_1-1) = q_2-U_2(d)(r_2-1) = 0$,
$q_1-U_1(d)(r_1-1) > 0 < q_2-U_2(d)(r_2-1)$ and
\begin{align}
  \label{eq:ComparativeMixingConditionHr}
  H_{r_1}(\alpha) \geq (>) (\gamma^{r_1} +
  (1-\gamma)^{r_1})H_{r_1}(\beta) + H_{r_1}(\gamma) \nonumber\\
  \Rightarrow
  H_{r_2}(\alpha) \geq (>) (\gamma^{r_2} +
  (1-\gamma)^{r_2})H_{r_2}(\beta) + H_{r_2}(\gamma)
\end{align}
or $q_1-U_1(d)(r_1-1) < 0 > q_2-U_2(d)(r_2-1)$ and
\eqref{eq:ComparativeMixingConditionHr} holds with opposite inequalities.

Define the \textcite{renyi_measures_1961} entropy $R_r(\alpha) = \ln(\alpha^r +
(1-\alpha)^r)/(1-r)$. Notice that $R_r(\alpha) \geq R_r(\beta)$ if and only if $H_r(\alpha) \geq H_r(\beta)$
and $H_r(\alpha) \geq (\gamma^r + (1-\gamma)^r) H_r(\beta) + H_r(\gamma)$ if and only if $R_r(\alpha) \geq R_r(\beta) +
R_r(\gamma)$. That is, the \textcite{renyi_measures_1961} and
\textcite{tsallis_possible_1988} entropy are order-equivalent.

\begin{lemma}
  Suppose $0 \leq r < s$, $\alpha,\beta,\gamma \leq 1/2$, and
  \begin{align}
    \label{eq:EqualityR}
    R^r(\alpha) = R^r(\beta) + R^r(\gamma)
  \end{align}
  then,
  \begin{align}
    \label{eq:InequalityR}
    R_s(\alpha) > R_s(\beta) + R_s(\gamma).
  \end{align}
\end{lemma}
\begin{proof}
  It is straightforward to show that $\alpha\geq \beta$ and $\alpha \geq \gamma$ since for $r > 0$,
  $R^r(\gamma) \geq 0$. We substitute: $x_1 = \alpha^{r-1}$, $x_2 = (1-\alpha)^{r-1}$, $y_1=
  (\beta\gamma)^{r-1}$, $y_2=(\beta(1-\gamma))$, $y_3=((1-\beta)\gamma)^{r-1}$, $y_4=((1-\beta)(1-\gamma))^{r-1}$,
  $w_{ij}=(x_iy_j)^{1/(r-1)}$ and exponentiate both sides to obtain that
  \eqref{eq:InequalityR} is equivalent to:
  \begin{align}
    \label{eq:SubstitutedInequalityR}
    sgn(1-s) \sum_{ij} w_{ij}(x_i^t-y_j^t) > 0
  \end{align}
  where $t = (s-1)/(r-1)$.

  Note that the vector $y$ with weights $(w_{11}+w_{21}, \ldots ,w_{14}+w_{24})$ is a
  mean-preserving spread of the vector $x$ with weights $(w_{11}+\ldots+w_{14},
  w_{21}+\ldots+w_{24})$ since by \eqref{eq:EqualityR}, we have that
  \begin{align}
    \label{eq:SubstitutedEqualityR}
    \sum_{ij} w_{ij}(x_i-y_j) = 0
  \end{align}
  Since $y$ is a mean-preserving spread of $x$, we have by the properties of
  generalized means that $M^t(\vec{w},\vec{x}) \equiv (\sum_{ij} w_{ij}x_i^t)^{1/t} >
  (\sum_{ij}w_{ij}y_j^t)^{1/t} \equiv M^t(\vec{w},\vec{y})$ if $t < 1$ and the reverse
  inequality holds if $t > 1$. It follows that $\sum_{ij} w_{ij}(x_i^t-y_j^t)$ is
  negative if $t>1$ or $t<0$ and positive if $0<t<1$. Since $0<t<1$ holds if and
  only if $s<1$, \eqref{eq:SubstitutedInequalityR} holds.
  \end{proof}

  The lemma establishes that if at some $r$ we have that $(\alpha,\beta,\gamma)$ are such that
  $a \sim_1 b$, then at a higher $r$ it must be the case that
  $a \succ_2 b$.
  Since irrespective of the choice of $s$ the LHS of \eqref{eq:InequalityR} is
  increasing in $\alpha \leq 1$ and the RHS is increasing in $\beta \leq 1/2$, $\gamma \leq 1/2$,
  it
  follows that
  $\{(a,b): a \succsim_1 b\} \subseteq \{(a,b): a \succsim_2 b\}$

\section{Further Comparative Statics}\label{sec:furtherComparativeStatics}
The comparative statics of Section \ref{sec:comparativeStatics} focus with
respect to the parameter $q$ of the representation on the property whether
$U(a)(r-1)-q \gtreqless 0$. The present section provides additional results that require
the existence of nontrivial preference on a set of {\em consequences}, i.e.
elements of $\mathscr{S}$ that cannot be written in the form $\mu a \oplus (1-\mu) b$ with $\mu \in
(0,1)$. Let $\mathscr{X}$ be a set of consequences. The following definitions are the
standard definitions of certainty equivalents and comparative risk aversion for
mixture sets adjusted to the procedural mixture setting.
\begin{definition}[Certainty Equivalent]
  The certainty equivalent $c=ce(\alpha x \oplus (1-\alpha) y) \in \mathscr{X}$ of a procedural mixture of
  outcomes $x$ and $y$ is an outcome that fulfills $\alpha x \oplus (1-\alpha) y \sim \alpha c \oplus
  (1-\alpha) c$.
\end{definition}
\begin{definition}[Comparative Risk Aversion]
  $\succsim_1$ is at least as risk averse as $\succsim_2$ if for all $\alpha\in (0,1)$ and all $x,y,z
  \in \mathscr{X}$, we have that
  \begin{align}
    \label{eq:ComprativeRiskAversion}
    \alpha x \oplus (1-\alpha)y \succsim_1 & \alpha z \oplus (1-\alpha) z\\
    \Rightarrow \quad
    \alpha x \oplus (1-\alpha)y \succsim_2 & \alpha z \oplus (1-\alpha) z
  \end{align}
\end{definition}
Provided with our adjusted definitions, we can prove the following standard
result for decisions under risk which extends to the procedural case:
\begin{proposition}\label{prop:comparativeRisk}
  Let $\succsim_1$ and $\succsim_2$ be mixture entropy values with representations $U_1$ and
  $U_2$ and parameters $r_1$, $q_1$ and $r_2$, $q_2$, respectively. Let $U_1(\mathscr{X})$
  and $U_2(\mathscr{X})$ be convex sets. Suppose there exist some $x,y \in \mathscr{X}$ such that $x\succ_1 y$.
  Then the following statements are equivalent.
  \begin{enumerate}
  \item $\succsim_1$ is at least as risk averse as $\succsim_2$.
  \item The restriction of $U_1 $ to $\mathscr{X}$ is a concave monotone
    transformation of $U_2$ and $r_1 = r_2$.
  \end{enumerate}
\end{proposition}
\begin{proof}
  $\Leftarrow$ is trivial, we prove $\Rightarrow:$ It is straightforward to show that for all $z,w
  \in \mathscr{X}$, $z \succ_1 w$ if and only if $z \succ_2 w$. Since $H_r(1)=0$ and $\succsim_1$ and $\succsim_2$
  are continuous, it follows that utilities over outcomes must be continuous
  monotone transformations of another, i.e., $U_1 = T \circ U_2$ when restricted to
  $\mathscr{X}$.

  Since the values of outcomes are a convex set, for every $x,y \in \mathscr{X}$ such that
  $x \succ_2 y$ we can find $z$ such that $z = ce_1(1/2 x \oplus 1/2 y)$. Notice that by
  the definition of a certainty equivalent we have $U_1(x)/2 + U_1(y)/2 =
  U_1(z)$. If $T$ is not concave, then for some such $x$ and $y$, $ U_2(x) +
  U_2(y) = T^{-1}(U_1(x))/2 + T^{-1}(U_1(y))/2 < T^{-1}(U_1(z)) = U_2(z)$. But
  then $1/2 x \oplus 1/2 y \succsim_1 1/2 z $ but $1/2 x \oplus 1/2 y \prec_2 1/2 z \oplus 1/2 z$,
  contradicting that $\succsim_1$ is at least as risk averse as $\succsim_2$. We have thus
  established that $T$ is concave.

  Notice now that if $r_1 \neq r_2$, then $p_1(\alpha) \equiv \frac{\alpha^{r_1}}{\alpha^{r_1} +
    (1-\alpha)^{r_1}} \neq \frac{\alpha^{r_2}}{\alpha^{r_2} + (1-\alpha)^{r_2}} \equiv p_2(\alpha)$. Without loss
  of generality, assume that $p_1(\alpha) \geq p_2(\alpha)$ for $\alpha \geq 1/2$. Since $T$ is
  monotone and continuous, it is differentiable almost everywhere. Without loss
  of generality, assume $T$ is differentiable at $U_2(x)$ and $U_2(x) = U_1(x)$
  and $\partial T(U_2(x))/\partial U = 1$. Then we can find outcomes $x' \succ x \succ x''$ such that
  $p_1(\alpha) T(U_2(x')) + (1-p_1(\alpha)) T(U_2(x'')) > U_2(x) > p_2(\alpha) U_2(x') +
  (1-p_2(\alpha)) U_2(x'')$ contradicting that $\succsim_1$ is at least as risk averse as
  $\succsim_2$.
\end{proof}

Mixture entropy values which are equally risk averse can be compared by how much
the value of consequences is compared to the value of mixing.
\begin{definition}[Comparative Consequentialism]
  $\succsim_2$ is at least as consequentialist than $\succsim_1$ if for all $\alpha \in (0,1)$ and
  all $x,y \in \mathscr{X}$, we have that
  \begin{align}
    \label{eq:ComparativeCuriosity}
    \alpha x \oplus (1-\alpha) x \succsim_2 & y \\
    \Rightarrow \quad
    \alpha x \oplus (1-\alpha) x \succsim_1 & y
  \end{align}
\end{definition}
\begin{proposition}\label{prop:comparativeCuriosity}
  Let $\succsim_1$ and $\succsim_2$ be equally risk averse mixture entropy values with
  representations $U_1$ and $U_2$ and parameters $r_1$, $q_1$ and $r_2$, $q_2$,
  respectively. Let $U_1(\mathscr{X})$ and $U_2(\mathscr{X})$ be convex sets. Suppose there exist
  some $x,y \in \mathscr{X}$ such that $x\succ_1 y$. Then the following statements are
  equivalent.
  \begin{enumerate}
  \item $\succsim_2$ is at least as consequentialist as $\succsim_1$.
  \item There exist $s \in \mathbb{R}_{>0}$ and $t \in \mathbb{R}$ such that $U_2(x) = s U_1(x) + t$
    for all $x \in \mathscr{X}$ and $s q_1 + t \geq q_2$.
  \end{enumerate}
\end{proposition}
\begin{proof}
  $\Leftarrow$ is trivial, we prove $\Rightarrow$: If $\succsim_1$ and $\succsim_2$ are equally risk averse, then
  $r_1=r_2=r$ and $U_1$ is an affine transformation of $U_2$ on outcomes. Assume
  for the moment that $U_1 = U_2$ on outcomes. Notice that since $H_r(\alpha) \geq 0$
  for all $r$ and all $\alpha$, \eqref{eq:ComparativeCuriosity} holds if and only if
  $q_1 \geq q_2$. Since $U_2$ may be an affine transformation of $U_1$ on outcomes,
  by the uniqueness properties of $U$, the desired result follows.
\end{proof}
This provides a link between the parameters of our representation and the
intensity of the value of mixing relative to the value of consequences. Notice
that this comparison is only meaningful when $r_1 = r_2$ and the value of
consequences are cardinally comparable.

\section{Proof of Theorem \ref{thm:disjointMixture}}
\begin{proof}
  The intuition for the result is simply that similar to a procedural mixture
  set, a mixture set in which independence only applies to disjoint mixtures
  allows for $a \sim^* b$ and $\mu a \oplus (1-\mu) b \not\sim \mu a \oplus (1-\mu) a = a$ (since $a$
  and $a$ do not have disjoint support). The key difficulty is that the disjoint
  independence axiom might not apply to sufficiently many elements of the
  mixture set to restrict preferences to the desired
  representation.\footnote{For example, if there are only two outcomes, disjoint
    independence does not restrict preferences.}

  To prove the result, we first extend $\succsim^*$ from $\Delta \mathscr{X}$ to a set $\Delta \mathscr{X}^{\infty}$ which
  is a mixture set generated from finite mixtures of countably many copies of
  $\mathscr{X}$. The resulting relation is additively separable across the different
  copies of $\mathscr{X}$. $\Delta \mathscr{X}^{\infty}$ contains a subset that is isomorphic to a procedural
  mixture set and on which $\succsim^*$ fulfills the vNM axioms. We thus have by
  Theorem 1 the desired representation on the procedural mixture set. Because
  disjoint mixtures in $\mathscr{X}$ coincide with procedural mixtures in the procedural
  mixture set, the utility representation on $\Delta\mathscr{X}$ fulfills
  \eqref{eq:disjointEntropyRepresentation}. The details of these steps follow
  below.

  Let $\mathscr{X}^{\infty} = \sqcup_{i=0}^{\infty} \mathscr{X}$ be the disjoint union of countably many copies of
  $\mathscr{X}$. $x_i^k \in \mathscr{X}^{\infty}$ refers to the $k$th copy of $x_i \in \mathscr{X}$. Let $\Delta X^{\infty}$ be
  the mixture set generated from $\mathscr{X}^{\infty}$. A generic
  element of $\Delta \mathscr{X}^{\infty}$ can therefore be represented by
  $m=\{(x_i^k,\mu_i^k),\ldots,(x_j^l,\mu_j^l)\}$ such that $\sum_{i,k} \mu_i^k = 1$.
  Note that $\mathscr{M} \subset \Delta X^{\infty}$.

  Let $\mathscr{J}$ be a partition of the support such that there are three elements that
  each contain an essential pair of outcomes. By standard results
  \parencite[e.g.,]{wakker_additive_1989}, our axioms\footnote{Disjoint
    independence implies preference separability across disjoint subsets of the
    support. Continuity implies topological connectedness of each dimension.} together with
  the existence of three essential pairs of outcomes with mutually disjoint
  support guarantees that there exists an additive representation $U(m) =
  \sum_{J}u((x_i^1, \mu_i^1)_{i \in J})$ on $\Delta\mathscr{X}$ such that each component is normalized
  to zero if $\sum_{i \in J} \mu_i^1=0$. We can uniquely extend the relation $\succsim^*$ on
  $\mathscr{M}$ to $\Delta \mathscr{X}^{\infty}$ by summation of the utilities obtaining a utility
  representation unique up to affine transformations of the form $\sum_k \sum_{J \in
    \mathscr{J}}u((x_i^k, \mu_i^k)_{i \in J})$.

  Let $\mathscr{P}$ be the closure of $\mathscr{X}$ under an operator $\oplus$, i.e., the minimal set
  such that $\mathscr{X} \subset \mathscr{P}$ and for all $p,q \in \mathscr{P}$ and all $\mu \in [0,1]$, we have that $\mu p
  \oplus (1-\mu) q \in \mathscr{P}$. Similarly, let $\approx$ be the minimal relation such that the
  procedural mixture axioms are fulfilled. Notice that the quotient set $\mathscr{P}/\approx$ is
  also a procedural mixture set with $=$ being the equivalence relation.

  A generic element of $\mathscr{P}/\approx$ can be represented by a finite set
  $\{(x_1,\mu_1),\ldots,(x_n \mu_n)\}$ where all $x_i \in \mathscr{X}$ with $x_i = x_j$ permitted
  also for $i \neq j$. Thus, there is a natural mapping from $\mathscr{P}/\approx$ into $\Delta \mathscr{X}^{\infty}$
  which we denote $\phi: \mathscr{P}/\approx \rightarrow \Delta \mathscr{X}^{\infty}$.

  Let $\succsim$ be defined by $p \succsim q$ if $\phi(p) \succsim^* \phi(q)$. It is straightforward to see
  that $\succsim$ fulfills Weak Order and Continuity. Independence follows from the
  fact that $\succsim^*$ has an additively separable representation across different
  copies of $\mathscr{X}$ contained in $\mathscr{X}^{\infty}$ and fulfills disjoint independence.

  It follows that $\succsim^*$ on $\mathscr{P}/\approx$ has an entropy adjusted expected utility
  representation $V: \mathscr{P}/\approx \rightarrow \mathbb{R}.$ Notice that on $\phi(\mathscr{P}/\approx) \subset \Delta\mathscr{X}^{\infty}$ we have that $V
  \circ \phi^{-1}$ is additively separable across the partition of indexes $J$. It
  follows that $V \circ \phi^{-1}$ is an affine transformation of $U$. The desired
  functional form of disjoint mixtures when $U$ is restricted to $\Delta \mathscr{X}$ follows.
\end{proof}

\section{Proof of Corollary \ref{coro:nestedLuceHick}}
\begin{proof}
  The functional form of the decision probabilities follows from
  \textcite{kovac_behavioral_2022}, Theorem 1. Within each category $S
  \in \mathscr{S}$, Corollary \ref{coro:luceHick} yields that the
  decision times can be represented by a Tsallis entropy $H_{r_S}$.
  From category-wise continuity and disjoint independence follows that
  the overall decision time takes the functional form $\tau(D) =
  f(\{H_{r_S}(p_{D \cap S})\}_{S \in \mathscr{S}}; \{p_D(S)\}_{S \in
    \mathscr{S}})$. Across categories, we can thus represent each set
  by a probability distribution over $\mathscr{S}$ and the entropy
  value in each category. By our richness assumption and the
  probabilities following a nested Luce model, for any combination of
  values of the entropies $\{H_{r_S}(\cdot)\}$ and any arbitrary
  probability over $\mathscr{S}$ we can indeed find a set $D$ that
  generates these entropy values and probability over categories.
  Thus, we have a mixture set which we endow with the support $supp(D)
  = {S \in \mathscr{S} | p_D(S \cap D) > 0}$. It is straightforward to see
  that axiom \ref{axiom:disjointIndependence} holds for this supported
  mixture set. Theorem 2 then yields the representation after defining
  $t_S(D) = T^{-1}(f(H_{r_S}(p_{D \cap S}),0,\ldots,0;1,0,\ldots,0))$.
\end{proof}

\printbibliography
\end{document}